%% file: paper.tex
\documentclass[letterpaper,twocolumn,10pt]{article}
\usepackage{usenix2019_v3}
\usepackage{xurl}
\usepackage[english]{babel}
\usepackage{algorithm}
\usepackage{subcaption}
\usepackage{algpseudocode}
\usepackage{booktabs}
\usepackage{multirow}
\usepackage{xcolor}
\usepackage{dashrule}
\usepackage{tikz}
\usepackage{amsmath}
\usepackage{amssymb}
\usepackage{microtype}
\usepackage{enumitem}
\usepackage{xspace}
\usepackage{tabularx}
\usepackage{amsthm}

\newcommand{\sysname}{Rondo\xspace}
\newcommand{\cpname}{capacity probing\xspace}
\newcommand{\oodname}{out-of-order degree\xspace}
\newcommand{\Oodname}{Out-of-Order Degree\xspace}
\newcommand{\icname}{ideal controller\xspace}
\newcommand{\mcname}{multipath controller\xspace}

\newcommand{\para}[1]{\noindent\textbf{#1.}}
\usepackage{xcolor}

\theoremstyle{definition}
\newtheorem{proposition}{Proposition}

\definecolor{cFPCC}{HTML}{1F77B4}
\definecolor{cDCQCN}{HTML}{FF7F0E}
\definecolor{cSwift}{HTML}{2CA02C}
\definecolor{cHPCC}{HTML}{D62728}
\definecolor{cPower}{HTML}{9467BD}

\newcommand{\legendmarker}[2]{\tikz[baseline=-0.6ex]{\ifcase#2\relax
      \fill[fill=#1] (0,0) circle (2pt);\or
      \fill[fill=#1] (-2pt,-2pt) rectangle (2pt,2pt);\or
      \fill[fill=#1] (-2pt,-1.5pt) -- (0,2.5pt) -- (2pt,-1.5pt) -- cycle;\or
      \fill[fill=#1] (-2pt,1.5pt) -- (0,-2.5pt) -- (2pt,1.5pt) -- cycle;\or
      \fill[fill=#1] (2.2pt,0) -- (-1.6pt,2.2pt) -- (-1.6pt,-2.2pt) -- cycle;\fi
  }}

\begin{document}

\date{}

\title{\Large \bf \sysname: Controlling Reordering in High-Throughput Packet Spraying}

\author{
{\rm Xizhi Zhang\textsuperscript{1}\quad Zhongjie Chen\textsuperscript{2}\quad Yang Zhang\textsuperscript{1}\quad Fengyuan Ren\textsuperscript{1}}\\
\textsuperscript{1}Tsinghua University\qquad \textsuperscript{2}Microsoft Research
}
\maketitle

\begin{abstract}
\input{sections/abstract.tex}
\end{abstract}

\input{body}

\bibliographystyle{plain}
\bibliography{reference}

\section*{APPENDIX}
\appendix
\input{sections/appendix/appendix.tex}

\end{document}

%% file: sections/abstract.tex
Modern datacenter applications demand high network bandwidth, motivating packet-level multipath transport to exploit the aggregate capacity of multiple paths.
However, packet reordering under packet spraying poses challenges for hardware transports. Excessive packet reordering makes it difficult to infer packet loss, imposes substantial pressure on NIC resources, and ultimately degrades performance.
Existing transports either prioritize throughput without controlling packet reordering, or mitigate reordering by restricting transmissions, which leaves available path capacity unused.

In this paper, we identify two fundamental sources of packet reordering: transient inter-path queueing differences caused by control dynamics, and persistent path-delay differences due to path heterogeneity.
These observations motivate a joint design of a multipath controller and a packet scheduler to address transient and persistent delay differences, respectively.
We present \emph{\sysname}, a packet-level multipath transport that achieves both high throughput
and low packet reordering.
Its multipath controller uses ACK-based per-path capacity estimation and probing to maintain stable path loads, while its packet scheduler adjusts packet transmission order to compensate for persistent delay differences.
Testbed experiments and large-scale simulations show that \sysname reduces the maximum \oodname
by up to 97.5\% while maintaining high throughput.
With constrained receiver buffers, \sysname achieves up to 8.51$\times$ higher goodput.

%% file: body.tex
\input{sections/introduction/introduction_draft.tex}
\input{sections/motivation/motivation.tex}

\input{sections/analysis/analysis_revision.tex}
\input{sections/design/design_revision.tex}

\input{sections/implementation.tex}
\input{sections/evaluation/evaluation.tex}

\input{sections/conclusion.tex}

%% file: sections/introduction/introduction_draft.tex
\section{Introduction}
Hardware transport, such as Remote Direct Memory Access (RDMA), has become a key enabler of high-throughput, low-latency communication for datacenter applications. Datacenter networks usually employ multiple parallel paths~\cite{rfc7938,al-faresScalableCommodityData2008}.
Exploiting this path diversity for hardware transports~\cite{rfc7938,al-faresScalableCommodityData2008,qianAlibabaHPNData2024,araujoResilientAISupercomputer} can benefit throughput-oriented applications, ranging from distributed storage and data analytics to large language model (LLM) training and inference~\cite{qianAlibabaHPNData2024,renTENTDeclarativeSlice2026,wuDualPathAcceleratingAgentic2026}.

In recent years, packet-level multipath transport, or packet spraying, has gained increasing attention~\cite{shalevCloudOptimizedTransportProtocol2020,ultraethernetconsortiumUltraEtherneSpecification2025,singhviFalconReliableLow2025,araujoResilientAISupercomputer,dixitImpactPacketSpraying2013,leSTrackReliableMultipath2024,bonatoREPSRecycledEntropy2026a,luMultiPathTransportRDMA2018,qiSGLBScalableRobust2025}.
These transports select a path for each packet, allowing even a small number of flows to utilize capacity across multiple paths. Unfortunately, packet spraying can introduce packet reordering, posing particular challenges for hardware transports.

Gaps in the received packet sequence may indicate either packet loss or out-of-order arrival. RoCEv2, a single-path protocol that enforces strict in-order delivery, interprets any out-of-order arrival as packet loss. Even a small \oodname can cause severe performance degradation~\cite{luMultiPathTransportRDMA2018, liPSNPATHWhenMultipath2026}. Recent hardware transports are designed to tolerate limited packet reordering~\cite{shalevCloudOptimizedTransportProtocol2020,ultraethernetconsortiumUltraEtherneSpecification2025,singhviFalconReliableLow2025,araujoResilientAISupercomputer}, but still face two challenges. First, excessive packet reordering makes packet loss hard to infer. There is an inherent tradeoff between loss detection and out-of-order tolerance: a higher reordering tolerance delays loss recovery, whereas a lower one makes multipath transport more susceptible to spurious retransmissions. Second, excessive packet reordering imposes substantial pressure on NIC memory.
Tracking out-of-order packets requires metadata, while strict in-order delivery also requires reorder buffers for payload. Across many flows, both metadata and reorder buffers may exceed NICs' MB-scale on-chip memory budgets~\cite{luMultiPathTransportRDMA2018, mittalRevisitingNetworkSupport2018a,wangSRNICScalableArchitecture2023}.
Once the NIC exhausts its on-chip resources, it may have to drop incoming packets or apply backpressure to the sender. The resulting retransmissions and sender stalls reduce throughput and increase tail latency.

However, existing multipath transports struggle to keep packet reordering low. First, these transports have to make a tradeoff between bandwidth utilization and packet reordering. For example, Falcon~\cite{singhviFalconReliableLow2025} and UET~\cite{ultraethernetconsortiumUltraEtherneSpecification2025} use receiver backpressure to reduce the sending window when receiver resource occupancy is high. MP-RDMA~\cite{luMultiPathTransportRDMA2018} prunes slower paths that contribute to excessive reordering. Both strategies can leave available path capacity unused, sacrificing throughput to reduce reordering. Second, these transports control packet reordering reactively, responding only after it has occurred. Under bursty traffic, feedback delays cause reordering to accumulate before these mechanisms can respond.

We revisit the root causes of packet reordering to identify how multipath transports can proactively limit it.
We make two observations.
First, existing multipath transports introduce substantial inter-path queueing differences. During synchronized flow startup, these transports often inject large bursts before assessing path conditions, creating queue imbalances when traffic is unevenly distributed.
Even with a fixed set of active flows, continual per-path load adjustments cause queues to build and drain asynchronously, leading to fluctuating queueing delay differences across paths. Thus, an ideal multipath transport should both maintain shallow queues and balance load across paths.

Second, even when queues are stable, persistent delay differences across heterogeneous paths can still cause packet reordering.
Excluding slower paths can help but at the cost of bandwidth utilization~\cite{luMultiPathTransportRDMA2018}. Our key insight is that once queueing delays stabilize, packet scheduling can compensate for persistent delay differences across paths. For example, the multipath transport can send packets with higher packet sequence numbers (PSNs) earlier on higher-delay paths, followed by lower-PSN packets on lower-delay paths.

Based on these observations, we present \sysname, a packet-level multipath transport that achieves both high throughput and low packet reordering.
\sysname coordinates the multipath controller and a packet scheduler to jointly achieve these goals.
The multipath controller combines ACK-rate-based congestion control with deficit-based path selection to stabilize per-path loads and sustain high utilization. It also uses a low-priority capacity probing mechanism to accelerate flow startup with minimal impact on queueing.
With queueing delays stabilized by the multipath controller, the packet scheduler adjusts transmission order to compensate for persistent delay differences across paths, eliminating packet reordering.

In summary, we make the following contributions.

\begin{itemize}[nosep]
\item  Analyzing the impact of reordering on multipath transport, and identifying two sources of packet reordering: inter-path queueing differences and persistent path-delay differences that require different remedies.
\item Designing \sysname, a packet-level multipath transport that achieves high throughput with low and controllable packet reordering.
\item Implementing \sysname in DPDK and evaluating it through extensive experiments. In typical datacenter workloads, \sysname eliminates out-of-order arrivals under incast
and reduces maximum \oodname by 28.6\%--97.5\%
under the other traffic patterns,
while achieving JCT comparable to or lower than that of state-of-the-art multipath transports. With a reorder buffer of approximately one BDP, \sysname achieves 1.15$\times$--3.60$\times$ the goodput
of other multipath transports, increasing to
1.38$\times$--8.51$\times$ as the buffer becomes more scarce.
\end{itemize}

%% file: sections/motivation/motivation.tex
\section{Background and Motivation}
\label{sec:background}

\subsection{Multipath in Datacenters}
\label{sec:bg-whyps}

\input{sections/motivation/whyps.tex}

\subsection{Reordering Issues}
\label{sec:bg-ood}

\input{sections/motivation/disorderissue.tex}

%% file: sections/motivation/whyps.tex
Modern datacenter topologies are typically provisioned with multiple paths between racks. Exploiting the path diversity can aggregate their available bandwidth, avoid congested paths, and tolerate path failures.
Since RDMA is a single-path transport by design with a strict in-order delivery requirement, the most common scheme is flow-level load balancing, i.e., equal-cost multipath (ECMP). However, it has been widely studied to suffer from hash collisions and fail to balance traffic effectively~\cite{qianAlibabaHPNData2024,luMultiPathTransportRDMA2018}. Flowlet-level schemes improve adaptability by allowing a flow to switch paths after an inter-packet gap~\cite{alizadehCONGADistributedCongestionaware2014a,qureshiPLBCongestionSignals2022,vaniniLetItFlow2017}. With a sufficiently large gap, a flowlet can change paths while largely preserving packet order. However, high-throughput transfers rarely expose gaps for flowlet formation, and this scheme essentially prevents the opportunity to aggregate multiple paths' bandwidth~\cite{huFLBFinegrainedLoad2025}.

Recent multipath transports use a much more fine-grained approach, i.e., packet spraying, to distribute packets of the same flow across multiple paths~\cite{shalevCloudOptimizedTransportProtocol2020,ultraethernetconsortiumUltraEtherneSpecification2025,singhviFalconReliableLow2025,araujoResilientAISupercomputer}. A key component of a packet spraying transport is the \emph{multipath controller}, which consists of  congestion control and path selection. The congestion control regulates sending rates at either the flow or path granularity, while the path selection distributes packets across paths based on different rules. Although packet spraying enables fine-grained load balancing and efficient utilization of available paths, it can easily cause substantial packet reordering. Unfortunately, it's challenging for hardware transports to handle packet reordering effectively~\cite{luMultiPathTransportRDMA2018}.

%% file: sections/motivation/disorderissue.tex
In this section, we examine the costs of packet reordering. We introduce two metrics to quantify the impact: \emph{\Oodname} (OOD), the difference between the largest PSN received and
the next expected PSN; \emph{Reorder Buffer Occupancy} (RBO), the number of out-of-order packets buffered while the receiver waits for missing earlier packets.
OOD is always no smaller than RBO, and
the two metrics usually follow the same trend in practice.

\para{Ambiguity between packet loss and reordering}
When the received packet sequence contains a gap, the transport cannot immediately determine whether the missing packet was lost or merely reordered.
Even in lossless networks with PFC enabled, packet loss remains
unavoidable in production due to link failures and faulty devices~\cite{qianAlibabaHPNData2024,gangidi2024rdma}.
RoCEv2 treats all gaps as packet losses and triggers inefficient Go-Back-N retransmissions~\cite{luMultiPathTransportRDMA2018,liPSNPATHWhenMultipath2026}. Selective Repeat (SR), adopted by the latest ConnectX NICs, still generates a NAK for each out-of-order packet, degrading performance significantly under heavy packet reordering~\cite{singhviFalconReliableLow2025}. Falcon and UET instead introduce RACK-TLP and OOD-based Early Recovery (ER) to distinguish packet loss from reordering~\cite{ultraethernetconsortiumUltraEtherneSpecification2025,singhviFalconReliableLow2025}. However, they still face a fundamental tradeoff: recovering losses promptly risks spurious retransmissions when packet reordering is severe.

\begin{figure}[!t]
    \centering
    \captionsetup[subfigure]{skip=0pt,belowskip=0pt}
    \begin{subfigure}[t]{0.522\linewidth}
        \centering
        \includegraphics[width=\linewidth]{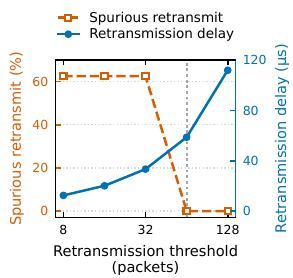}
        \caption{}
        \label{fig:moti_retx}
    \end{subfigure}
    \hfill
    \begin{subfigure}[t]{0.452\linewidth}
    \centering
    \includegraphics[width=\linewidth]{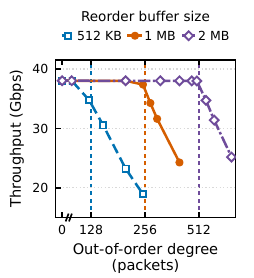}
    \caption{}
    \label{fig:moti_rbsize}
    \end{subfigure}
    \caption{(a) There is an inherent tradeoff between loss detection and out-of-order tolerance. (b) Throughput drops sharply as OOD approaches the reorder-buffer limit.}
\end{figure}

To demonstrate this trade-off, we implement ER in a DPDK-based multipath transport and configure the switch to induce an OOD of 64 packets and a random packet drop rate of 0.1\%.
ER retransmits an unacknowledged packet when the gap between its PSN and the highest acknowledged PSN exceeds a fixed threshold.
Figure~\ref{fig:moti_retx} shows that suppressing spurious retransmissions under this reordering requires higher thresholds, at the cost of delayed loss recovery.

\para{Reordering poses memory pressure}
Many datacenter applications rely on the strict in-order delivery semantics provided by the transport layer. To absorb out-of-order packets, the receiver must buffer them until the missing earlier packets arrive. However, commodity NICs typically provide only several megabytes of on-chip memory~\cite{wangSRNICScalableArchitecture2023}, making it impractical to provision a dedicated reordering buffer for every flow. Falcon, for example, chooses to use a shared one-BDP reorder buffer across all flows~\cite{singhviFalconReliableLow2025}. When RBO exhausts this capacity, the receiver must drop packets, leading to performance degradation, head-of-line blocking or even deadlock~\cite{jiangOrderlockNewType2025}.
Falcon mitigates such buffer pressure by adjusting sending windows based on receiver buffer occupancy.

We implement a similar mechanism in DPDK and measure it on our ConnectX-7 NICs.
The switch is configured to create a delay difference across paths, corresponding to a target OOD under line-rate transmission.
As shown in Figure~\ref{fig:moti_rbsize}, throughput begins to degrade once OOD becomes comparable to the reorder-buffer capacity, and decreases further as OOD grows.
A large OOD increases receiver buffer pressure, causing the sender to reduce its window and preventing it from sustaining line rate.

\para{Direct Data Placement (DDP) is not a panacea}
Some transports~\cite{ultraethernetconsortiumUltraEtherneSpecification2025,araujoResilientAISupercomputer,wangSRNICScalableArchitecture2023,shalevCloudOptimizedTransportProtocol2020} use DDP to write packets directly to their final memory locations regardless of arrival order. However, this technology is only applicable to some AI/ML workloads, whereas many datacenter applications rely on in-order delivery semantics~\cite{luMultiPathTransportRDMA2018,singhviFalconReliableLow2025}. Even emerging AI/ML applications such as DeepEP~\cite{deepep} require ordering guarantees across consecutive operations and therefore cannot run over transports that do not support in-order delivery~\cite{ucclep, fabriclib}. In addition, DDP does not fully eliminate the costs of packet reordering: packet loss remains difficult to distinguish from reordering, and tracking out-of-order packets still consumes NIC resources even without reorder buffers.
For example, IRN~\cite{mittalRevisitingNetworkSupport2018a}, a single-path, DDP-based transport, reserves five BDP-sized bitmaps for each QP. This per-QP metadata scales linearly with the number of QPs and can therefore consume substantial NIC memory~\cite{wangSRNICScalableArchitecture2023}. We confirmed this limitation on ConnectX-7 NICs, which also support DDP; Appendix~\ref{app:hw_cx7} provides details.

\para{Summary}
Excessive packet reordering degrades transport performance by obscuring packet loss and increasing receiver-side resource pressure.
Controlling reordering can therefore improve transport efficiency, reduce buffer occupancy,
and even benefit DDP-based transports that do not require a reordering buffer.

%% file: sections/analysis/analysis_revision.tex
\begin{figure*}[t]
    \centering
    \begin{minipage}[t]{0.35\linewidth}
        \centering
        \includegraphics[width=\linewidth]{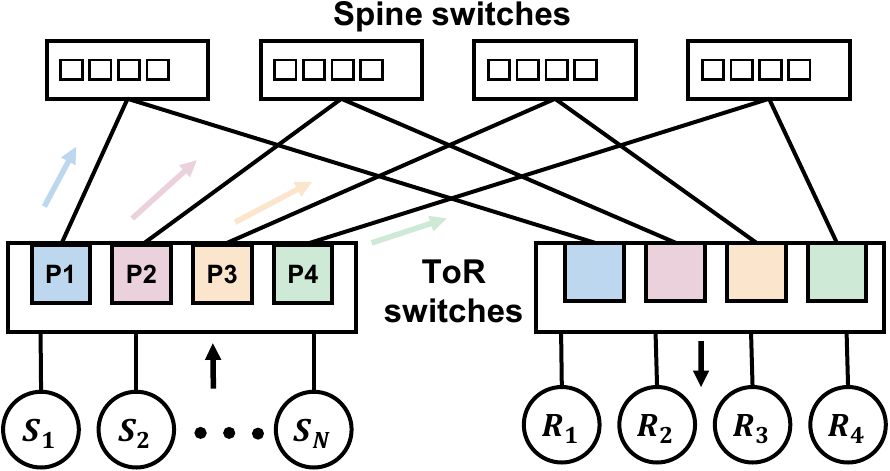}
        \caption{Multipath topology with four parallel paths.
}
        \label{fig:ana_topo}
    \end{minipage}
    \hfill
    \begin{minipage}[t]{0.63\linewidth}
        \centering
        \includegraphics[width=\linewidth]{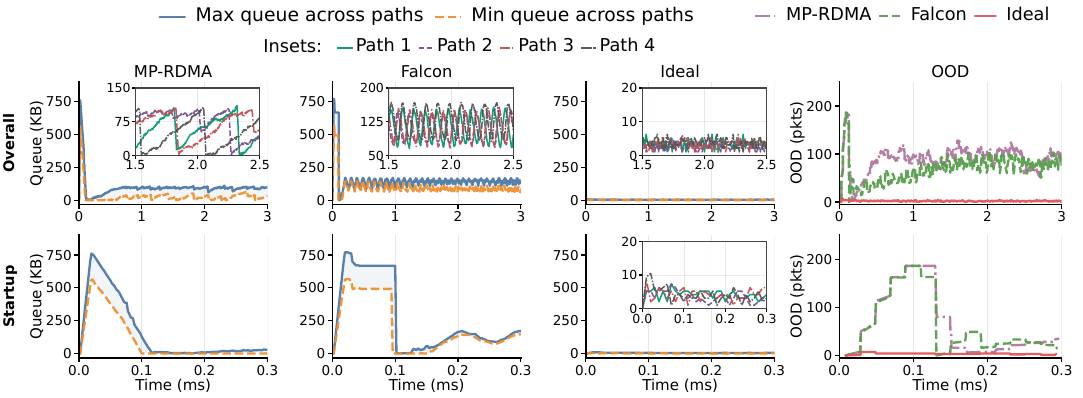}
        \caption{Queue dynamics and NIC-level OOD under synchronized flows.  NIC-level OOD denotes the sum of the OOD of all flows at the NIC.}
        \label{fig:ana_symm}
    \end{minipage}
\end{figure*}

\section{Understanding the Sources of Packet Reordering}
\label{sec:understanding}

In this section, we first analyze how inter-path delay differences translate into packet reordering, and then examine how these delay differences arise in multipath transport.

\subsection{What Determines Out-of-order Degree?}
\label{sec:oom}

We denote the forward delay of path $p$ at time $t$ as $D_{p}(t)$.
Consider packet $i$ transmitted at time $t_i$ on path $p_i$. It will arrive at the receiver at time $t_i+D_{p_i}(t_i)$.
We say a packet $j$ \textit{overtakes} packet $i$ if it is transmitted later but arrives earlier:
\begin{align}
    t_j > t_i
    \quad\text{and}\quad
    t_j+D_{p_j}(t_j)
    <
    t_i+D_{p_i}(t_i),
    \label{eq:overtake_i}
\end{align}

The PSN gap ($j-i$) is the number of packets transmitted during the overtaking window ($t_j - t_i$), which can be approximated by the integral of the aggregate sending rate $R(u)$ over that window:
\begin{align}
    j - i
    \approx
    \int_{t_i}^{t_j} R(u)\,du
    \label{eq:psn_gap}
\end{align}
Suppose packet $i$ is the next expected packet at the receiver,
and $j$ is the highest-PSN packet that overtakes it.
Immediately before packet $i$ arrives, $j-i$ is exactly the
OOD.

Equation~\eqref{eq:overtake_i} shows that the overtaking window is bounded by the inter-path delay difference.
In high-throughput packet spraying, where inter-packet transmission gaps are small, the maximum overtaking window is approximately the maximum delay difference among all active paths.
Let $\bar R$ denote the average aggregate sending rate over this
window.  Equation~\eqref{eq:psn_gap} gives the following estimate of
the maximum OOD:
\begin{align}
    OOD_{\max}
    \approx
    \bar R
    \left(
        \max_{p\in\mathcal{P}} D_p(t)
        -
        \min_{p\in\mathcal{P}} D_p(t)
    \right)
    \label{eq:ood_estimation}
\end{align}
Appendix~\ref{app:general-ood-analysis} provides the detailed derivation.

Equation~\eqref{eq:ood_estimation} shows that the maximum OOD is determined mainly by two factors: the aggregate sending rate and the inter-path delay difference.
To maximize $\bar R$ near the available network capacity while keeping the OOD under control, the inter-path delay difference should be minimized.

However, existing multipath transports do not perform ideally in this regard, which, as we will show in the following two subsections, is due to two different sources of inter-path delay difference: (1) imbalanced queuing delay and (2) persistent path-delay heterogeneity.

\subsection{Inter-path Queuing Delay Imbalance}
\label{sec:cid}

We decompose the delay of path $p$ into a persistent path-specific delay $\bar{D}_p$ and a queuing delay $\tilde{D}_p(t)$:
\begin{align}
    D_p(t)=\bar{D}_p+\tilde{D}_p(t)
\end{align}
On homogeneous paths, $\bar{D}_p$ is identical across paths, so inter-path delay differences arise primarily from $\tilde{D}_p(t)$.
Existing multipath transports keep each queue shallow, but do not necessarily keep queuing delays balanced across paths.

We demonstrate this problem through ns-3~\cite{NS3NetworkSimulator} using the topology in Figure~\ref{fig:ana_topo}.
Sixteen senders simultaneously transmit to four receivers over four parallel paths bottlenecked at ToR ports P1--P4.
We compare two representative multipath transports, MP-RDMA~\cite{luMultiPathTransportRDMA2018} and Falcon~\cite{singhviFalconReliableLow2025}, with a transport using an ideal controller.
The ideal controller has perfect prior knowledge of all flows and path states, and evenly divides the bottleneck bandwidth among the competing flows to determine each flow's ideal capacity share.
It then controls each flow's per-path inflight according to this capacity share, such
that the aggregate inflight on each path remains at BDP plus 5\,KB.
Figure~\ref{fig:ana_symm} shows the result.

\para{Capacity-agnostic rate ramp-up causes severe transient imbalance}
Within the first 0.1~ms after flow startup, MP-RDMA and Falcon exhibit inter-path queue-length differences of $\sim$200~KB, causing OOD to peak at approximately 200 packets.
 At startup, paths may differ in both residual bandwidth and the number of newly arriving flows. Existing multipath controllers inject traffic aggressively before learning each path's available capacity, overloading different paths to different
degrees and thereby creating transient inter-path queue imbalance.
The same problem arises when a transport rapidly increases its sending rate to utilize spare bandwidth.

\para{Reactive load adjustment across paths causes sustained oscillations}
After 1.5~ms, MP-RDMA and Falcon enter a steady phase, yet their queues differ by $\sim$100~KB, resulting in an OOD of approximately 100 packets. The congestion signals used by these transports, ECN and delay, reflect the congestion level on each path, but do not directly indicate the load that each flow should maintain on that path. Therefore, multipath controllers continuously adjust each path's load in response to congestion feedback. These reactive adjustments occur asynchronously across paths, causing queues to build and drain at different times and creating sustained queue oscillations, as shown in the insets of Figure~\ref{fig:ana_symm}.

\para{Ideal queue balancing effectively tames such reordering}
The \icname perfectly balances load across paths and fully utilizes available bandwidth, exhibiting almost zero OOD.
It shows that excessive packet reordering is not an inherent consequence of packet spray.
However, the foreseen and global knowledge is impractical because feedback is delayed and incomplete, and control is distributed across flows.
Taming reordering while keeping high throughput remains still challenging.

\subsection{Persistent Path-delay Heterogeneity}
\label{sec:hid}

Equal-cost physical paths may not have identical $\bar{D}_p$ in practice.
Paths may span different physical distances or traverse heterogeneous switching equipment, while link failures can further exacerbate path asymmetry.
In production datacenters, these effects can introduce persistent inter-path delay differences ranging from several to more than ten microseconds~\cite{gangidi2024rdma,bernardi2026rngflatdatacenternetworks,f10vincent}.
Unlike queuing-delay imbalance, balancing load across paths cannot eliminate reordering caused by persistent path-delay heterogeneity.

We demonstrate this problem using the topology in Figure~\ref{fig:ana_topo}, with four senders, each transmitting to a different receiver.
We vary the propagation delay of a link on path~4 from 2\,$\mu$s to 27\,$\mu$s in 5\,$\mu$s increments, creating a persistent delay difference of up to 25\,$\mu$s between path~4 and the other paths.

\begin{figure}[!t]
    \centering
    \includegraphics[width=0.65\linewidth]{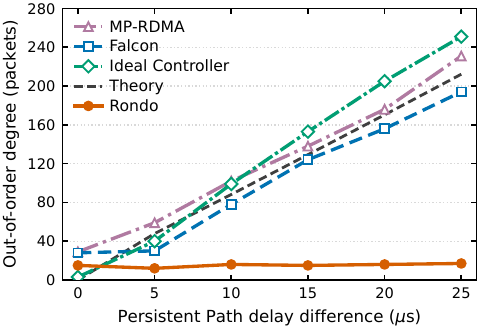}
    \caption{The OOD under persistent delay difference.}
    \label{fig:ana_unsymm}
\end{figure}

\para{Persistent path-delay heterogeneity causes reordering that cannot be eliminated by controllers}
Figure~\ref{fig:ana_unsymm} shows that OOD increases with the persistent inter-path delay difference under MP-RDMA, Falcon, and even the \icname.
A persistent path-delay difference of 15\,$\mu$s can cause a NIC-level OOD of 124--153 packets.

What's more, as the persistent path-delay difference grows and becomes the dominant
source of inter-path delay difference, it enlarges the overtaking window
and thus increases OOD.
This trend is consistent with Equation~\eqref{eq:ood_estimation}, which provides a linear estimation of OOD based on the inter-path delay difference and the aggregate sending rate, as shown in the theoretical line in Figure~\ref{fig:ana_unsymm}.

A naive way to reduce such OOD is assigning less traffic
to a slower path or avoiding it entirely.
However, doing so leaves the path bandwidth underutilized.
The desired transport behavior is to utilize the available transmission opportunities
on slower paths while mitigating the resulting OOD.

%% file: sections/design/design_revision.tex
\section{\sysname Design}
\subsection{Overview}

The observations in \S\ref{sec:understanding} lead to two design insights
for controlling OOD without sacrificing throughput.
First, to suppress reordering caused by inter-path queueing imbalance,
the \mcname should approach the behavior of the ideal
controller in \S\ref{sec:cid}, which matches per-path load to available capacity precisely and avoids large queue oscillations.
Second, persistent path-delay heterogeneity requires mechanisms beyond load control.
That motivates introducing a \emph{packet scheduler} to the transport, which adjusts the packet transmission order to
compensate for persistent inter-path delay heterogeneity.

Based on the above analysis, we propose \emph{\sysname},
a multipath transport that co-designs the \mcname and the packet scheduler.
The \mcname aims for reducing queue fluctuations, and also maintains stable
path-delay conditions for the packet scheduler.
By transmitting packets assigned to slower paths earlier, the packet scheduler
can better align their arrival order with packets sent on faster paths.
It fully utilizes the transmission opportunities provided by the controller, so it preserves the per-path load established by the controller.

\begin{figure}[!t]
    \centering
    \includegraphics[width=1.0\linewidth]{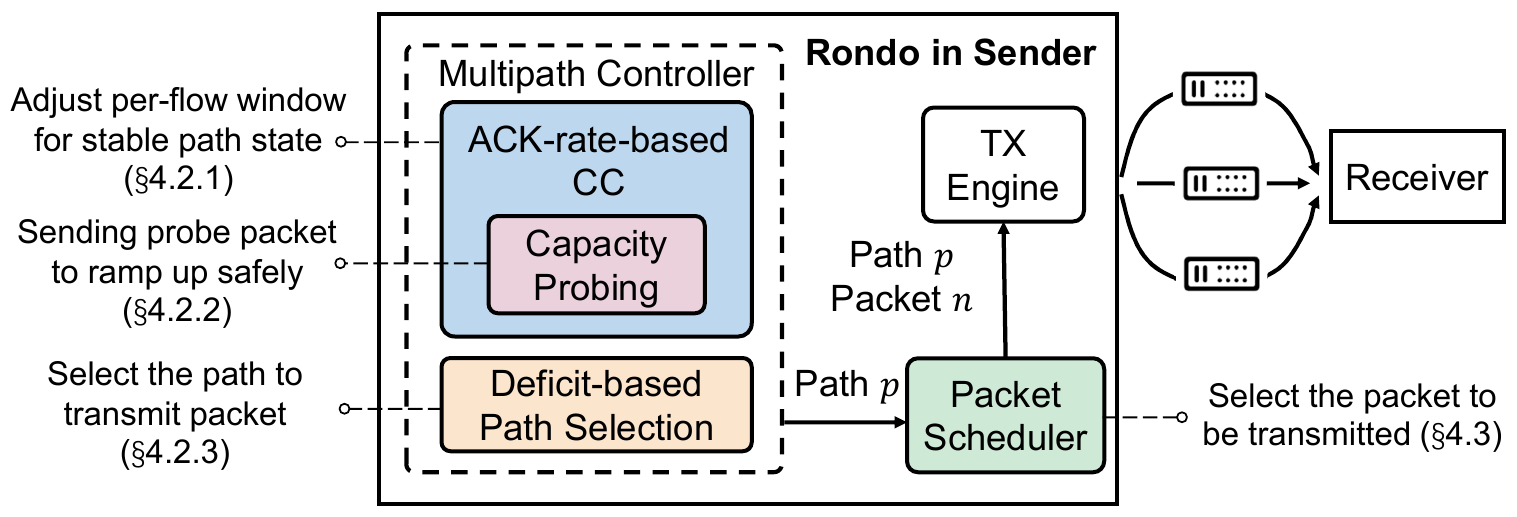}
    \caption{The architecture of \sysname.}
    \label{fig:overviews}
\end{figure}

Figure~\ref{fig:overviews} shows the architecture of \sysname.
The \mcname determines when the sender can transmit and
which path to use.
Given a transmission selected by the controller, the packet
scheduler selects the packet to transmit on that path.
The receiver returns per-packet ACKs carrying the corresponding PSN and
path information, and buffers out-of-order packets until they can be delivered
in order.

\subsection{Multipath Controller}
The multipath controller in \sysname should achieve the precise and stable
load control demonstrated by the ideal controller.
It consists of three components.
ACK-rate-based congestion control (ARCC) adjusts the per-path windows,
capacity probing rapidly and safely discovers newly available bandwidth,
and deficit-based path selection smoothly distributes transmissions
according to the per-path windows.

\subsubsection{ACK-rate-based Congestion Control}
\label{sec:arcc}
Maintaining stable per-path load requires a quantitative target for each flow's load on that path.
Existing congestion signals indicate congestion but do not provide such a target, causing queue oscillations.
The ideal controller uses global knowledge to calculate the appropriate allocation of bottleneck capacity among flows,
but such knowledge is unavailable in practice.
ARCC instead uses per-path ACK rate to estimate the capacity share currently obtained by each flow.

\para{ACK rate as a capacity indicator}
At a fully utilized bottleneck with bandwidth $C$, the capacity shares
currently obtained by all flows traversing it sum to $C$.
Each sender can observe its own share through its ACK rate on this path.
ARCC uses this observation to construct a target per-path window
\begin{align}
    W^*_p = \hat r_p T + q_p,
    \label{eq:arcc_update}
\end{align}
where $\hat r_p$ is the measured ACK rate of this flow on path $p$,
$T$ is the base RTT, and $q_p$ is a small per-flow margin.

\begin{figure}[!t]
    \centering
    \includegraphics[width=0.75\linewidth]{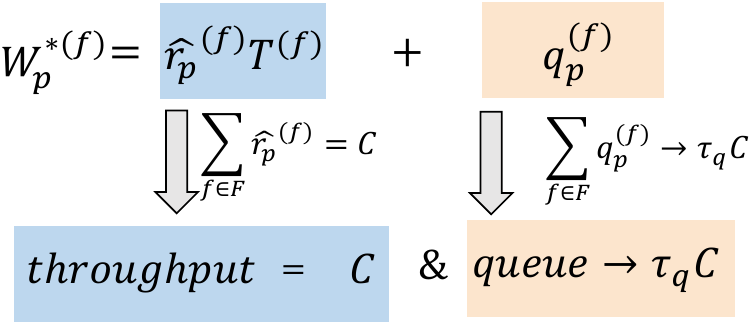}
\caption{Rationale of ARCC.}
    \label{fig:arcc}
\end{figure}

Figure~\ref{fig:arcc} illustrates the rationale of this target window.
The first term $\hat r_pT$ constitutes most of the window according to the capacity share observed from the ACK rate.
$\hat r_p T$ of all flows collectively provides just enough inflight
to sustain the bottleneck capacity.
The margin $q_p$ provides a small amount of additional inflight,
allowing flows to adjust their capacity shares while controlling the queue.
ARCC adjusts $q_p$ using AIMD such that the sum of $q_p$ across flows approaches $\tau_q C$, where $\tau_q$ is
the target queueing delay.
Because throughput is primarily sustained by the ACK-rate term, only
the small $q_p$ term requires reactive adjustment and can therefore be
updated conservatively.
This allows ARCC to maintain high utilization with a shallow and stable
queue.

\para{Algorithm}
ARCC updates per-path windows every $T_u$.
At each update on path $p$, ARCC computes the average ACK rate over the last interval
and then applies an Exponentially Weighted Moving Average (EWMA) to obtain an ACK rate estimate $\hat r_p$.
This progress filters out short-term fluctuations in ACK arrivals.
ARCC uses a method similar to Swift~\cite{kumarSwiftDelaySimple2020} to obtain the base RTT $T$,
which can be directly derived from the datacenter topology.
It then estimates the pipe size as $C_p \gets \hat r_p T$ and the
flow's excess inflight as
$Q_p \gets pktInfl-C_p$, which serves as an estimate of
its queued data.

ARCC next performs AIMD on $q_p$ to drive the queueing delay toward the target $\tau_q$.
When the estimated queue is below the target, it additively increases
$q_p$ up to $q_0$; otherwise, it multiplicatively decreases $q_p$
according to the ratio between the target and estimated queue.
Finally, ARCC computes $W_p^*$ according to Equation~\eqref{eq:arcc_update}.
If the flow's estimated queue is large,
ARCC directly
sets $W_p$ to $W_p^*$ to quickly drain the queue.
Otherwise, it applies an EWMA to smooth the window update.

\para{Per-path window with flow-level pacing}
Pacing smooths packet transmission and helps maintain a stable path load.
However, maintaining an independent pacer for every path of
every flow increases the state overhead in NIC hardware~\cite{saeedCarouselScalableTraffic2017,radhakrishnanSENICScalableNIC}.
ARCC therefore uses a flow-level pacer.
After each window update, ARCC sums the per-path windows and computes
the aggregate pacing rate as
$\sum_{p} W_p/T$.
The pacing rate bounds the aggregate sending rate of the flow, while
 per-path windows independently control the per-path inflight.

\subsubsection{Capacity Probing}
\label{sec:bwprobe}

ARCC works well when the bottleneck is fully utilized,
but the ACK rate cannot reveal unused bandwidth when a path is underutilized.
A new mechanism is needed for $\mathcal{O}(1)$\footnote{
  We use $\mathcal{O}$ notation to characterize how the time required for
CC to converge to the efficient state scales with the deviation of the current state from the target state~\cite{zhangoscar26}.
} rate increase.
However, existing rate-increase mechanisms face a trade-off between fast
bandwidth discovery and path stability.
ECN and delay cannot quantitatively reveal the spare capacity in an underutilized link, so conservative
ramp-up takes $\mathcal{O}(BDP)$ or $\mathcal{O}(log(BDP))$ rounds~\cite{zhangoscar26}.
Aggressive packet-train-based approaches~\cite{andersonPCPEfficientEndpoint} converge faster but can build large transient
queues~\cite{agarwalFRCCProvablyFair2026}, while in-network telemetry requires additional switch support.

\para{Low-priority probing}
The key idea of \cpname is to translate the spare capacity into ACK streams without filling it with data packets.
\sysname uses dedicated, low-priority MTU-sized probe packets that carry no application payload.
The probe packets use a reserved PSN distinct from normal data packets.
During probing,
the sender transmits probe packets whenever data packet transmission is blocked by the per-path window.
The probes fill the remaining transmission opportunities up to line rate.

\begin{figure*}[!t]
    \centering
    \begin{minipage}[t]{0.33\textwidth}
        \centering
        \includegraphics[width=0.95\linewidth]{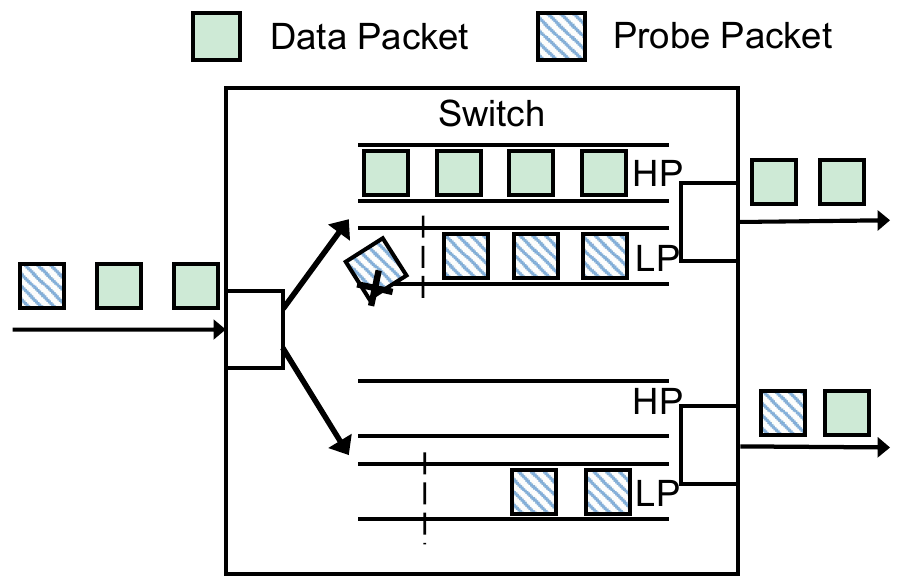}
        \caption{Probes use low-priority queues
        and consume only spare bandwidth.}
        \label{fig:bwprobe}
    \end{minipage}
    \hfill
    \begin{minipage}[t]{0.65\textwidth}
        \centering
        \includegraphics[width=\linewidth]{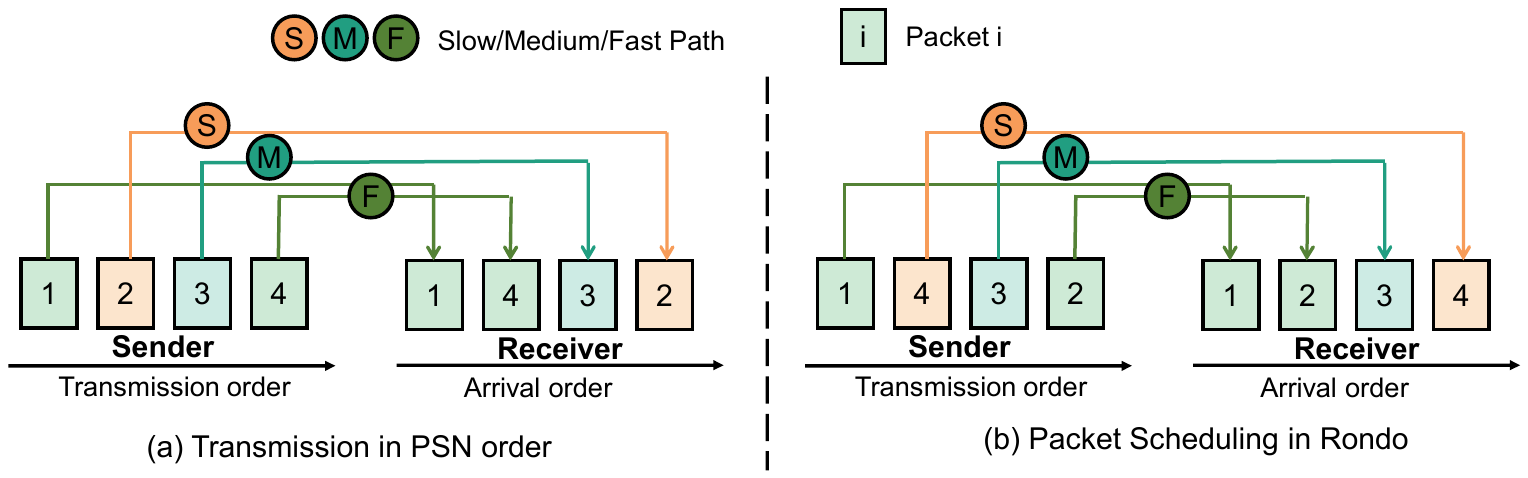}
        \caption{PSN-order transmission can cause out-of-order arrivals,
while \sysname adjusts the transmission order to mitigate it.}
        \label{fig:oosend}
    \end{minipage}
\end{figure*}

As shown in Figure~\ref{fig:bwprobe},
switches place probe packets in a low-priority probe queue, while normal data packets use the high-priority queue.
The switch applies strict-priority scheduling,
so it serves data packets first and transmits probe packets only with bandwidth left unused by data packets.
\sysname configures a small drop threshold for the probe queue, so excess probe packets are dropped to save the buffer.
Since probe packets carry no application payload, their loss does not require retransmission.
Both strict-priority queues and configurable per-queue drop thresholds are standard QoS mechanisms
supported by commodity datacenter switches~\cite{CiscoNexus9000}.
These mechanisms allow \sysname to aggressively probe for spare capacity,
while introducing little interference with data packet transmission and switch buffer space.

\para{Converting probe ACKs into data capacity}
The receiver acknowledges each arriving probe packet.
The sender distinguishes probe ACKs from normal data ACKs by their PSNs and
measures the probing ACK rates separately.
Because probe packets receive service
only from capacity left unused by data packets,
the probe ACK rate reveals the additional capacity that the path can provide to the flow.
\sysname incorporates this probe ACK rate
into ARCC's per-path capacity estimate and increases the data window accordingly.
To use the discovered capacity promptly, each probe ACK allows the sender to transmit one additional data packet before the first window update that incorporates probe ACKs.

\para{Triggering probing}
\sysname activates probing at flow startup and when a path appears to have spare capacity.
At flow startup, each path starts
with an initial window $W_0$, and probes for additional capacity.
During transmission, if all data packets sent at least one base RTT
earlier have been acknowledged for five consecutive updates, \sysname
starts a probing phase lasting three base RTTs.
Small increases in spare capacity can be captured by ARCC's margin term $q_p$, and \cpname is therefore primarily used for rapid ramp-up when substantial spare capacity becomes available.

\subsubsection{Deficit-based Path selection}
Per-path windows determine how much load each path can carry,
but the sender still needs to decide how to distribute individual transmissions across multiple available paths.
Round-robin or random selection~\cite{dixitImpactPacketSpraying2013,liPSNPATHWhenMultipath2026,EnablingPacketSpraying} does not account for differences in
path conditions, while selecting paths with larger open
windows~\cite{singhviFalconReliableLow2025} can create bursty packet assignments.
\sysname therefore uses deficit-based path selection to distribute
transmissions according to the per-path windows while avoiding bursts.

For each path $p$, \sysname computes its open window as $ O_p = \max(0, W_p-pktInfl_p)$.
When ARCC permits a transmission, \sysname increases the deficit $Z_p$ of every available path by its normalized open-window share,
\begin{align}
    Z_p \leftarrow Z_p + \frac{O_p}{\sum \limits_{j \in P} O_j}.
\end{align}
It then selects the path with the largest deficit and decreases its deficit by one.
The deficit tracks the difference between how many transmissions a path
should have received according to its open-window share and how many it
has actually received.
A path that has received less traffic accumulates deficit and is more likely to be selected subsequently.

\input{sections/design/oosend.tex}
\input{sections/design/design_theory.tex}

%% file: sections/design/oosend.tex
\subsection{Packet Scheduler}
\label{sec:ooosend}

Even with stable path control, persistent inter-path delay differences can cause
packets to arrive out of order.
Since the multipath controller has already determined the appropriate
per-path load, the scheduler should mitigate such reordering without
changing the transmission time or path selected by the controller.
As shown in Figure~\ref{fig:oosend}, existing transports always send the smallest unsent PSN for each transmission opportunity, and
the packet on a slower path may be overtaken by packets transmitted later on faster paths.
The packet scheduler in \sysname instead selects higher-PSN packets for transmissions on slower
paths, better aligning the resulting arrival order with the PSN order.

\para{Computing the skip}
Recall from \S\ref{sec:hid} that, for a transmission opportunity on path $p$,
we can estimate how many packets transmitted later on the other paths are
expected to arrive before the packet sent on path $p$.
\sysname sums these estimates and skips the corresponding number of unsent
packets, leaving the skipped lower-PSN packets for those earlier-arriving
transmissions.  For a transmission on path $p$,
\begin{align}
    \mathrm{skip}_p =
    \left\lfloor
    \sum_{\substack{p' \in \mathcal{P}\\p' \neq p}}
    \frac{W_{p'}}{\mathrm{RTT}_{p'}}
    \max\!\left(D_p-D_{p'},0\right)
    \right\rfloor
    \label{eq:packet-skip}
\end{align}
where $D_{p'}$ is the measured forward
delay of path $p'$.
For a packet sent on path $p$, packets transmitted on a faster path $p'$
within $D_p-D_{p'}$ can arrive earlier.
Multiplying this interval by the transmission rate on $p'$ estimates their
number.
The sum therefore estimates the current transmission opportunity's relative
position in the expected arrival order across all paths.
\sysname transmits the packet with the $(\mathrm{skip}_p+1)$-th smallest
unsent PSN, leaving the preceding lower-PSN packets for transmissions
expected to arrive earlier.

\para{Delay measurement}
Packet scheduling requires the forward-delay difference across paths.
For each used path, \sysname periodically timestamps a packet to measure the forward delay and RTT.
The scheduler does not require fine-grained clock synchronization because
the sender-receiver clock offset cancels when computing inter-path delay
differences.

The skip calculation assumes stable inter-path delay differences and sufficiently
smooth short-term transmission rates.
The former is provided by the controller's stable queue control,
while flow-level pacing and deficit-based path selection suppress
bursty packet assignments.
Moreover, \sysname enables packet scheduling only after the forward-delay
estimate remains stable for several consecutive samples.

%% file: sections/design/design_theory.tex
\subsection{Theoretical Analysis}
\label{sec:theory_analysis}
We analyze \sysname with a fluid model, summarizing the conclusion below and providing proofs in Appendix~\ref{app:theory_analysis}.

\begin{proposition}[Fairness]
\label{prop:arcc_fairness}
For $N$ flows' logical paths\footnote{A logical path is a transport-level path identifier that the network maps to a physical forwarding path.} passing a single bottleneck
of capacity $C$, ARCC converges geometrically to equal throughput $C/N$.
\end{proposition}

\begin{proposition}[Out-of-order degree bound]
\label{prop:ood_bound}
\sysname provides the following worst-case bound on the OOD caused by
inter-path queue differences:
\begin{equation}
    OOD \le
    \frac{R_{\max}}{C}
    \frac{T_{\max}}{T_{\min}}
    \left(C\tau_q + \frac{N_l q_0}{4} + N_S W_0\right)
\end{equation}
where $N_l$ and $N_S$ are the upper bounds on active and
 starting logical paths in the bottleneck, respectively;
 $R_{\max}$ is the maximum aggregate sending rate among these flows;
$T_{\min}$ and $T_{\max}$ are the minimum and maximum base RTTs.
This is a conservative worst-case bound that requires extreme queue
imbalance across paths.
\end{proposition}

\begin{proposition}[Packet scheduling correctness]
\label{prop:delay_compensation}
Suppose each path
\(p\) of a flow has an arbitrary but fixed forward delay \(D_p\) and a constant
sending rate \(r_p\).
If the delays and rates are accurately measured, the packet
scheduler eliminates the OOD caused by persistent path-delay
heterogeneity.
\end{proposition}

%% file: sections/implementation.tex
\section{Implementation}
\label{sec:impl}
We implement a \sysname prototype in DPDK~\cite{dpdk}
with $\sim$3000 LoC.
Using DPDK's mlx5 PMD, the prototype follows a
datapath similar to that of an RNIC.
All \sysname control logic runs on the sender's DPDK lcore.
We implement UDP-source-port-based path steering, strict-priority
scheduling, and tail drop in $\sim$400 lines of P4 code.

\para{Overhead}
The DPDK implementation saturates the NIC line rate on our testbed.
We further estimate \sysname's arithmetic overhead based on~\cite{OverviewSifive7seriesMicroarchitecture}.
Path selection and packet scheduling are performed per packet,
while ARCC  and skip calculations run periodically off the critical path.
With four paths per flow, the estimated overhead is 25 cycles per packet
on the critical path and 170--210 cycles per flow
every $T_u$ for periodic control.
For comparison, single-path HPCC takes 180 cycles per ACK
under the same model~\cite{jiangOdinRethinkingCongestion2026}.
Appendix~\ref{app:imple_detail} provides implementation details,
optimizations, and the overhead breakdown.

\para{Compatibility}
\sysname is compatible with the transport semantics
and upper-layer interfaces of existing datacenter transports,
and relies on widely supported ACK feedback and delay measurement~\cite{kumarSwiftDelaySimple2020,jarrett-weeksInfiniBandSpecificationFrequently,ScalingZeroTouch2021}.
Packet scheduling may require additional support in
hardware transports, but Tonic~\cite{arashloo2020enabling} and
SR in modern RNICs suggest that flexible packet
selection is feasible in high-speed NICs.
The switch-side mechanisms require only standard strict-priority scheduling and configurable
queue drop thresholds available on commodity switches.

%% file: sections/evaluation/evaluation.tex
\section{Evaluation}

\begin{figure*}[!t]
    \centering
    \begin{subfigure}[t]{0.20\linewidth}
        \centering
        \includegraphics[width=\linewidth]{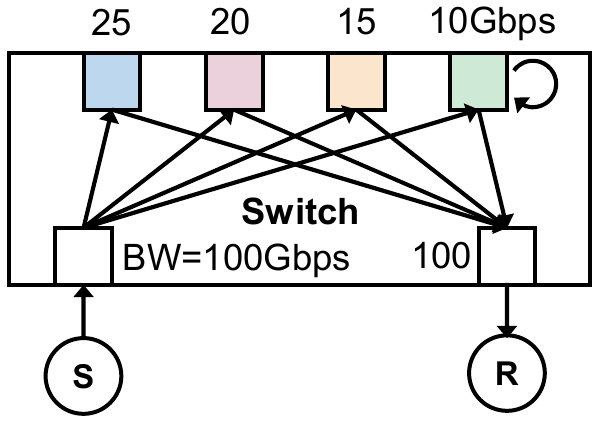}
        \caption{Testbed with four paths.}
        \label{fig:testbed_topo}
    \end{subfigure}
    \hfill
    \begin{subfigure}[t]{0.26\linewidth}
        \centering
        \includegraphics[width=\linewidth]{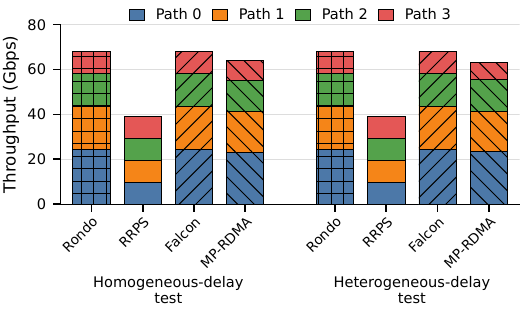}
                \captionsetup{
    margin=5pt,
    singlelinecheck=false
}
        \caption{\sysname fully utilizes the available path capacity.}
        \label{fig:testbed_tput}
    \end{subfigure}
    \hfill
    \begin{subfigure}[t]{0.26\linewidth}
        \centering
        \includegraphics[width=\linewidth]{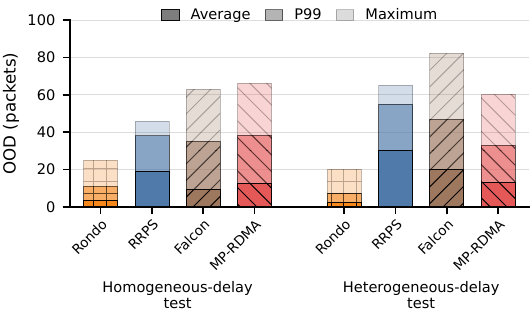}
                \captionsetup{
    margin=5pt,
    singlelinecheck=false
}
        \caption{\sysname limits the \oodname.}
        \label{fig:testbed_ood}
    \end{subfigure}
    \hfill
    \begin{subfigure}[t]{0.26\linewidth}
        \centering
        \includegraphics[width=\linewidth]{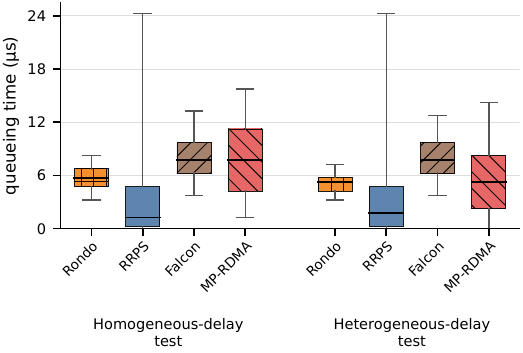}
                \captionsetup{
    margin=5pt,
    singlelinecheck=false
}
        \caption{\sysname reduces queueing-delay oscillations.}
        \label{fig:testbed_queueing}
    \end{subfigure}
    \caption{Testbed topology and results.}
    \label{fig:eval_testbed}
\end{figure*}

We evaluate \sysname using our DPDK prototype on a hardware
testbed and large-scale ns-3~\cite{NS3NetworkSimulator} simulations.
Our evaluation answers the following questions:
\begin{itemize}[leftmargin=*,itemsep=1pt,topsep=2pt,parsep=0pt]
    \item Can \sysname achieve high throughput while maintaining low OOD on a hardware testbed?
    (\S\ref{sec:testbed-result})

    \item How does \sysname compare with existing transports in throughput and OOD across diverse workloads and network conditions?
    (\S\ref{sec:simulation-result}--\S\ref{sec:heterogeneous-networks})

    \item How does reducing OOD improve performance?
    (\S\ref{sec:reordering-benefits})

    \item How does each design component contribute to \sysname's performance?
    (\S\ref{sec:microbenchmarks-and-ablation-study})
\end{itemize}

\para{Baselines}
We compare \sysname with ECMP and five representative multipath transports.
\begin{itemize}[leftmargin=*,itemsep=1pt,topsep=2pt,parsep=0pt]
    \item \textbf{ECMP}~\cite{hoppsAnalysisEqualCostMultiPath2000} + \textbf{DCQCN}~\cite{zhuCongestionControlLargeScale2015}:
    This configuration is widely deployed in today's RDMA datacenters.
    Since ECMP pins each flow to a single path,
    it does not introduce packet reordering in the absence of packet loss. It uses selective repeat for recovery.
    \item \textbf{MP-RDMA}~\cite{luMultiPathTransportRDMA2018}:
    MP-RDMA uses a per-flow, ECN-driven congestion control.
    Its ACK-clocked path selection transmits packets over responding paths and
    temporarily prunes paths whose delayed ACKs indicate excessive OOD.
    \item \textbf{Falcon}~\cite{singhviFalconReliableLow2025}:
    Falcon uses a per-path Swift-like~\cite{kumarSwiftDelaySimple2020} congestion control, adjusting its window
    according to the path's RTT. It transmits packets on the path with the maximum available window.
    It uses PLB~\cite{qureshiPLBCongestionSignals2022} to remap
    persistently congested paths, a NIC congestion window to avoid buffer exhaustion and per-path RACK-TLP for loss recovery.
    \item \textbf{UET}~\cite{ultraethernetconsortiumUltraEtherneSpecification2025}: UET uses NSCC,
    a per-flow congestion control that adjusts its window using both ECN and RTT.
    Its REPS path selection~\cite{bonatoREPSRecycledEntropy2026a} learns path
    conditions from ACKs and caches paths for packet spraying. UET uses ER and RTO with dynamic thresholds for loss recovery.
    \item \textbf{RoCE Adaptive Routing (AR)}~\cite{khashabHighspeedNetworkingGigaScale2026}:
    RoCE AR delegates path
    selection to switches, which choose output ports based on local queue occupancy.
    \item \textbf{Round-robin Packet Spraying (RRPS)}~\cite{EnablingPacketSpraying,liPSNPATHWhenMultipath2026}:   RRPS performs congestion-oblivious packet spraying by
    assigning packets to paths in a round-robin manner.
\end{itemize}

\subsection{Testbed Evaluation}
\label{sec:testbed-result}

\para{Testbed setup}
Our testbed uses two servers with ConnectX-7 NICs connected through
an Intel Tofino programmable switch over 100\,Gbps links.
We configure the switch to forward packets from the ingress port through one of four internal ports,
creating four equal-cost paths, as shown in Figure~\ref{fig:testbed_topo}.
In the homogeneous-delay test, all paths have the same base RTT of \textasciitilde20~$\mu$s, and we limit their bandwidths to 25, 20, 15, and 10~Gbps to emulate different degrees of congestion.
In the heterogeneous-delay test, we add an additional 10~$\mu$s delay to one path by chaining additional loopback stages.
We compare \sysname with RRPS, MP-RDMA and Falcon in the testbed.

\para{Result}
As shown in Figure~\ref{fig:testbed_tput},
\sysname achieves
68.18 and 68.14~Gbps in the homogeneous- and heterogeneous-delay
tests, respectively, matching Falcon's throughput while outperforming
RRPS and MP-RDMA.
Its per-path windows and deficit-based path selection adapt the
inflight traffic to each path's capacity.
In contrast, RRPS distributes packets evenly across paths,
so the 10-Gbps bottleneck path limits its aggregate throughput.
As shown in Figure~\ref{fig:testbed_ood}, \sysname reduces average OOD by 82--92\%
relative to RRPS, 64--88\% relative to Falcon, and 72--81\%
relative to MP-RDMA.
In the homogeneous-delay test,
this benefit mainly comes from better queue control, as shown in Figure~\ref{fig:testbed_queueing}.
In the heterogeneous-delay test,
packet scheduling further compensates for the persistent delay difference.

\subsection{Large-Scale Simulation}
\subsubsection{Simulation Setup}
\label{sec:setup}
We compare \sysname with ECMP, MP-RDMA, Falcon, UET and AR in the simulation. We exclude comparison with RRPS here as our testbed experiments show that it substantially underutilizes the available path capacity.

Except in \S\ref{sec:reordering-benefits}, all schemes use a sufficiently large reorder buffer to independently characterize each scheme's bandwidth utilization and OOD behavior, without conflating them with OOD-induced performance degradation. Specifically, each NIC has a 1.2~MB shared reorder buffer (approximately $8\times$ the BDP), which accommodates the entire observed OOD distribution. When the buffer is full, the NIC accepts only the next expected packet,
drops other packets, and triggers NAKs.

\para{Parameters}
We configure \sysname with
$T_u = 0.5\,base\_RTT$,
$q_0 = 3\,MTU$,
$\tau_q = 0.04\,max\_base\_RTT$,
the ACK-rate EWMA smoothing parameter $\alpha = 0.8$,
and $\gamma = 0.5$.
For the other baselines, we use the recommended parameter settings
when available and otherwise use the best-performing configuration
obtained through parameter tuning.

\para{Simulation topology}
We use a three-tier Clos topology~\cite{rfc7938} comprising 320 servers
across 5 pods. Each pod contains 4 ToR and 4 aggregation switches,
and 16 core switches interconnect the pods.
Server-to-ToR links operate at 100~Gbps, while inter-switch links operate at 400~Gbps.
The default topology is non-oversubscribed (1:1) and provides 4 equal-cost paths
between ToRs in the same pod and 16 between ToRs in different pods.
We also evaluate a 4:1 oversubscribed configuration with only 4 active core switches.

\para{Workloads}
We evaluate \sysname using the following workloads:
(i)~\emph{Typical traffic patterns} comprise incast, permutation,
and MoE-like all-to-all traffic.
Incast induces last-hop congestion, while permutation traffic has each host
send to one host and receive from exactly one host, stressing the network fabric.
In MoE-like all-to-all traffic, each host simultaneously sends data
to multiple randomly selected receivers.
(ii)~\emph{Trace-driven workloads} include both realistic datacenter traffic
and distributed training.
The datacenter workloads are derived from production Web Search~\cite{alizadeh2010data}
and Cache Follower~\cite{roy2015inside} traces.
For distributed training, we use STAGE~\cite{man2026stage} and
Astra-Sim~\cite{wonASTRAsim20ModelingHierarchical2023} to simulate the
communication trace of training a 7B MoE Transformer model.

\begin{figure*}
    \centering
    \includegraphics[width=\linewidth]{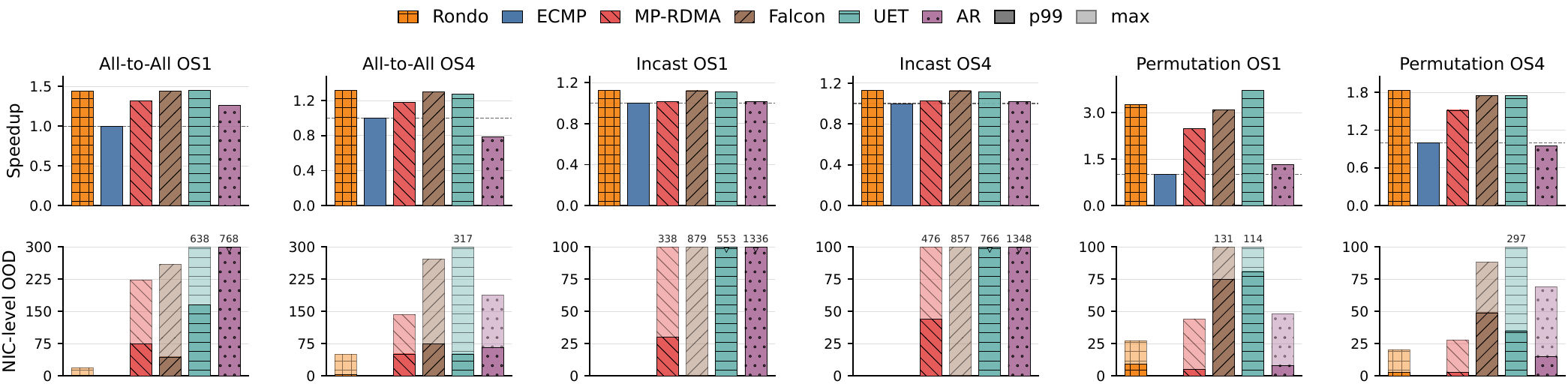}
    \caption{Performance under typical datacenter traffic patterns. JCT speedup is measured relative to ECMP.}
    \label{fig:classical_flow_performance}
\end{figure*}

\begin{figure}[t]
    \centering
    \includegraphics[width=\linewidth]{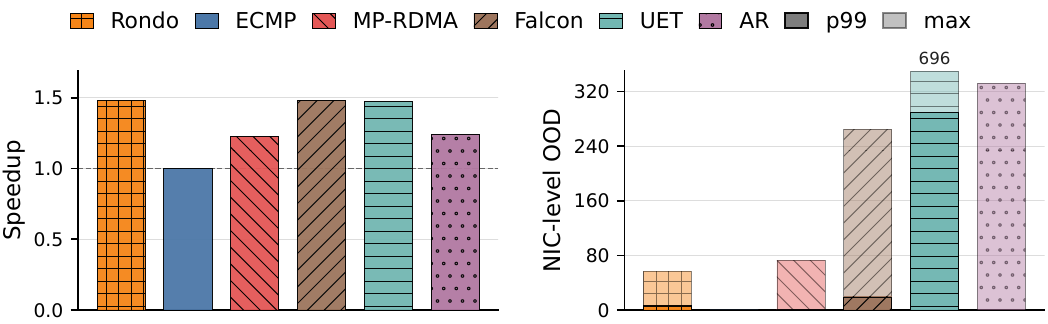}
    \caption{Communication-time speedup and NIC-level OOD during MoE Transformer training.
}
    \label{fig:eval_train}
\end{figure}

\subsubsection{Typical Traffic Patterns}
\label{sec:simulation-result}
Figure~\ref{fig:classical_flow_performance} shows the performance of different algorithms under typical datacenter traffic patterns
with 1:1 and 4:1 oversubscription. The NIC-level OOD is the sum of the OOD across all flows at the NIC.
In the all-to-all and permutation scenarios,
\sysname reduces the maximum NIC-level OOD by 28.6\%--91.5\% compared to MP-RDMA,
77.3\%--92.7\% compared to Falcon,
76.3\%--97.0\% compared to UET, and
43.8\%--97.5\% compared to AR.
In the incast scenario, \sysname does not cause any out-of-order arrivals,
whereas other multipath schemes exhibit maximum OOD values of 338--1348 packets.
Meanwhile, \sysname achieves speedups of 1.12$\times$--3.26$\times$
and provides the lowest JCT in four of the six scenarios,
with Falcon and UET showing comparable performance.
AR performs poorly under 4:1 oversubscription because it cannot capture end-to-end congestion.
\sysname's OOD reduction mostly comes from the controller's ability to keep queues stable,
while the throughput benefit comes from its rapid convergence to the
ideal operating point. In summary, \sysname substantially reduces OOD while preserving high throughput under all traffic patterns.

\subsubsection{Trace-Driven Workloads}
\label{sec:trace-driven-workloads}
\para{Distributed-training workloads}
We use Astra-sim to simulate the communication traffic of distributed ML workloads, including Allreduce, All-to-All, Allgather, Reduce-Scatter, and point-to-point communication.
Figure~\ref{fig:eval_train} shows that \sysname accelerates training communication by 1.48$\times$, matching Falcon and outperforming the other schemes.
Moreover, it controls both tail and maximum OOD, keeping them significantly lower than those of other multipath transports.

\begin{figure}[!t]
       \centering
    \begin{minipage}[t]{0.49\linewidth}
        \centering
        \includegraphics[width=\linewidth]{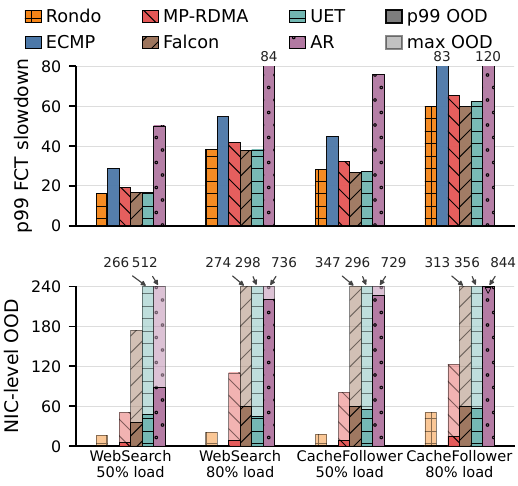}
        \caption{
\sysname improves tail FCT while reducing OOD under realistic workloads.
        }
        \label{fig:eval_dctrace}
    \end{minipage}
    \hfill
    \begin{minipage}[t]{0.49\linewidth}
    \centering
    \includegraphics[width=\linewidth]{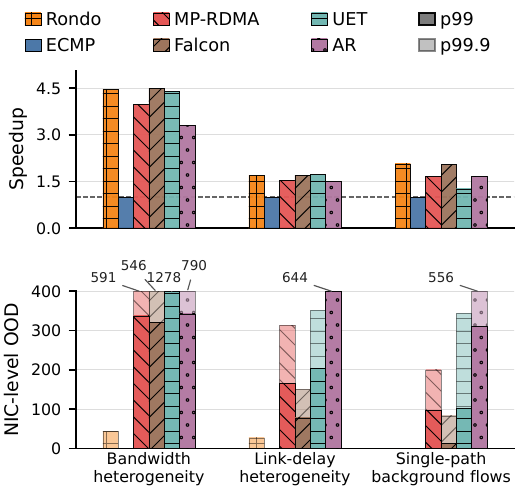}
    \caption{\sysname's performance under different heterogeneous scenarios.}
    \label{fig:eval_hete}
    \end{minipage}
\end{figure}

\para{Realistic datacenter workloads}
Figure~\ref{fig:eval_dctrace} shows that \sysname significantly reduces tail flow completion time (FCT) \footnote{FCT slowdown is defined as the ratio of a
flow's actual FCT to its ideal FCT under line-rate transmission.} under Web Search and Cache Follower workloads across different loads.
This result shows that multipathing also benefits general datacenter workloads.
These workloads contain more short flows and are harder to control than ML workloads,
but \sysname still guarantees a low OOD.

\subsubsection{Heterogeneous Network Conditions}
\label{sec:heterogeneous-networks}
We evaluate these schemes under different heterogeneous scenarios,
with the results shown in Figure~\ref{fig:eval_hete}.
For bandwidth heterogeneity, we randomly select 10\% of the links and reduce their bandwidth to one-quarter of the original capacity 50~$\mu$s after flows start.
For link-delay heterogeneity, we add 20~$\mu$s to the delays of randomly selected links.
We also add single-path background flows to the network for mixed traffic scenarios.

Across all heterogeneous scenarios, \sysname achieves the best speedup and maintains the lowest OOD. This is because, for heterogeneity caused by fabric bandwidth changes or background traffic, its ACK-rate measurements detect capacity changes and quickly adapt to the new capacities. For link-delay heterogeneity, its packet scheduler compensates for the delay differences to keep OOD low.

\begin{figure*}[t]
    \centering
    \begin{minipage}[t]{0.211\linewidth}
        \centering
        \includegraphics[width=\linewidth]{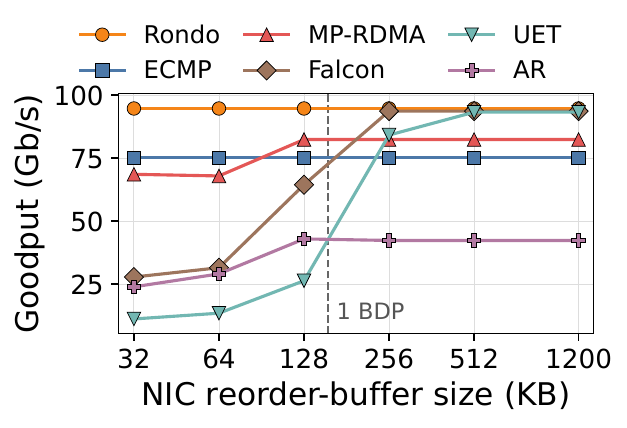}
        \captionsetup{
    margin=5pt,
    singlelinecheck=false
}
        \caption{
Aggregate goodput with different reorder-buffer sizes.
        }
        \label{fig:eval_buffersweep}
    \end{minipage}
    \hfill
    \begin{minipage}[t]{0.235\linewidth}
        \centering
        \includegraphics[width=\linewidth]{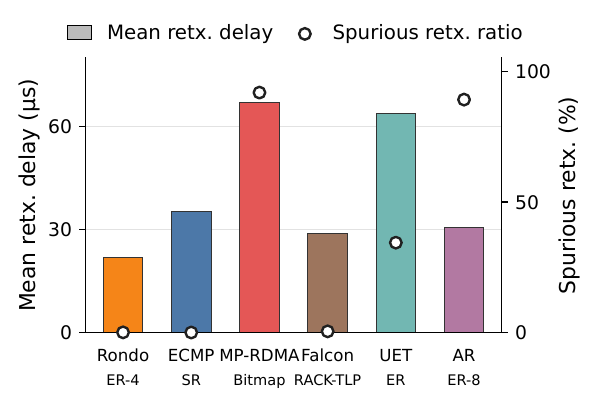}
        \captionsetup{
    margin=5pt,
    singlelinecheck=false
}
        \caption{Loss-recovery performance under 0.1\% random packet loss.}
        \label{fig:eval_spruretransratio}
    \end{minipage}
    \hfill
        \begin{minipage}[t]{0.235\linewidth}
    \centering
    \includegraphics[width=\linewidth]{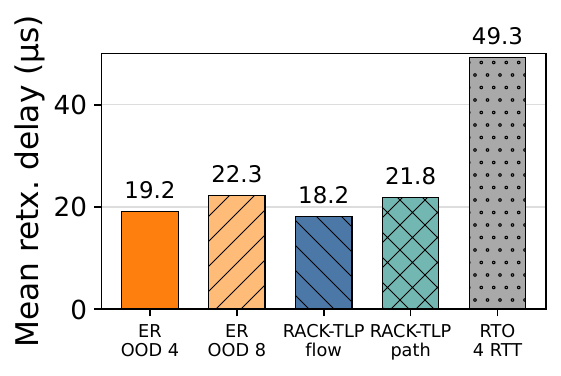}
    \captionsetup{
    margin=5pt,
    singlelinecheck=false
}
    \caption{Recovery delay with different recovery mechanisms in \sysname.}
    \label{fig:eval_differentretrans}
    \end{minipage}
    \hfill
    \begin{minipage}[t]{0.305\linewidth}
        \centering
        \includegraphics[width=\linewidth]{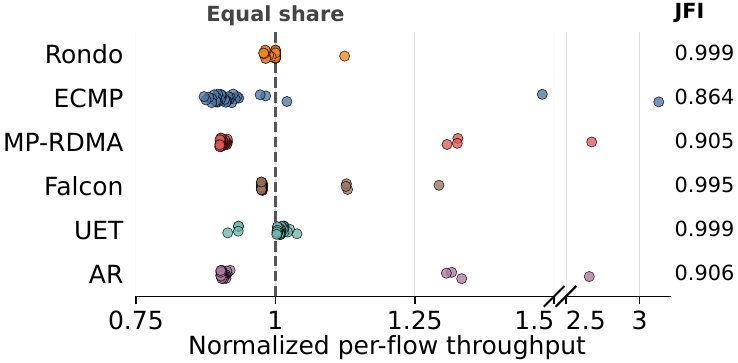}
        \captionsetup{
    margin=5pt,
    singlelinecheck=false
}
        \caption{Per-flow throughput distribution under incast, normalized by each scheme's mean throughput.}
        \label{fig:eval_fairness}
    \end{minipage}
\end{figure*}

\subsubsection{Benefits of Controlled Reordering}
\label{sec:reordering-benefits}
The preceding experiments deliberately exclude the performance degradation caused by limited reorder buffers to study each scheme's throughput and OOD behavior.
We next show that controlling OOD reduces reorder buffer requirements and enables prompt loss recovery without excessive spurious retransmissions.

\para{Controlled OOD sustains throughput with small reorder buffers}
Provisioning $O(\text{BDP})$ reorder buffer capacity in hardware is increasingly costly as link bandwidth grows.
We therefore vary the shared reorder buffer size from 1.2~MB to 32~KB under 4-to-1 incast with background flows.
Figure~\ref{fig:eval_buffersweep} shows that \sysname sustains 94.8~Gbps of goodput even with a 32~KB buffer because its bounded OOD rarely exhausts the buffer capacity.
In contrast, the other multipath transports generate more out-of-order arrivals; as the buffer shrinks, buffer exhaustion causes more packet drops.
With a buffer of approximately one BDP (used in Falcon\footnote{A NIC equipped with DRAM, such as Falcon, can extend its reorder buffer into on-NIC DRAM at the cost of higher access latency.}),
\sysname achieves 1.15$\times$--3.60$\times$ the goodput
of the other multipath transports.
With a 32~KB buffer, this advantage grows to
1.38$\times$--8.51$\times$.

\para{Controlled OOD enables fast and accurate loss detection}
A receiver cannot immediately determine whether a PSN gap is caused by packet loss or reordering.
Figure~\ref{fig:eval_spruretransratio} compares all schemes with their corresponding reliability mechanisms under a 0.1\% random packet-drop rate. \sysname with ER, ECMP, and Falcon with per-path RACK-TLP incur few spurious retransmissions, whereas the other schemes exhibit substantial spurious retransmission ratios.
\sysname also achieves a lower retransmission delay by bounding OOD, which enables prompt early retransmission.

\sysname does not prescribe a particular loss recovery mechanism and is
compatible with most existing recovery schemes.
Figure~\ref{fig:eval_differentretrans} further compares two other loss-recovery mechanisms in \sysname.
Early recovery with a four-packet threshold recovers losses in 19.2~$\mu$s while incurring only 0.05\% spurious retransmissions.
Flow-level RACK-TLP achieves a similar recovery delay of 18.2~$\mu$s but incurs 1.2\% spurious retransmissions.
Thus, controlling OOD allows \sysname to use a small reordering threshold for prompt loss recovery without sacrificing detection accuracy.

\subsection{Deep Dive into \sysname}
\label{sec:microbenchmarks-and-ablation-study}

\para{\sysname provides fairness for flows sharing the same bottleneck}
Figure~\ref{fig:eval_fairness} shows that \sysname provides similar per-flow throughput,
with a Jain fairness index of 0.999 under 32-to-1 incast.
For multipath transports in this incast scenario, the last-hop switch is the only bottleneck.
Under ECMP, however, flow collisions create other bottlenecks even when
some paths remain underutilized.

\begin{figure}[!t]
    \centering
    \begin{minipage}[t]{0.49\linewidth}
        \centering
        \includegraphics[width=\linewidth]{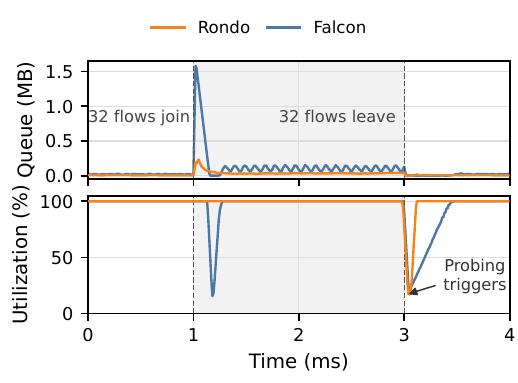}
        \caption[\sysname converges quickly after network changes.]{\sysname converges quickly.\protect\footnotemark}
        \label{fig:eval_react}
    \end{minipage}
    \hfill
    \begin{minipage}[t]{0.49\linewidth}
        \centering
        \includegraphics[width=\linewidth]{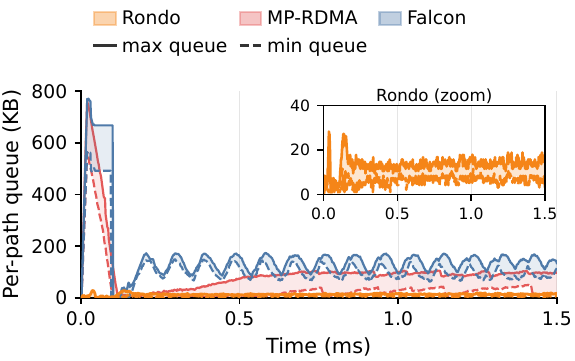}
        \caption{\sysname maintains stable queues.}
        \label{fig:eval_stable_queues}
    \end{minipage}
\end{figure}
\footnotetext{Utilization in this figure accounts only for data packets and excludes \sysname's probe packets.}

\para{The controller achieves fast convergence}
Figure~\ref{fig:eval_react} shows how \sysname reacts to flows joining and leaving.
Initially, 8 flows share the bottleneck.
When 32 flows join the bottleneck, \sysname transmits only $W_0$ data packets for each flow and uses probe packets to discover the remaining capacity, avoiding a large queue buildup.
ARCC then adjusts its window based on the ACK rate and drains the queue quickly without
reducing bandwidth utilization.
After these flows leave, \sysname triggers capacity probing and fully utilizes the
bandwidth in one step without overshooting.
These fast and precise reactions enable \sysname to achieve high performance.

\para{\sysname maintains stable queues}
We evaluate \sysname under the same setup as in \S\ref{sec:cid}
and present the result in Figure~\ref{fig:eval_stable_queues}.
The \sysname controller maintains low and stable queues,
keeping the inter-path queue difference always below 30~KB,
whereas the difference under other schemes can reach approximately 200~KB.

\begin{figure*}[t]
    \centering
    \begin{minipage}[t]{0.235\textwidth}
        \vspace{0pt}
        \centering
        \includegraphics[
            width=\linewidth
        ]{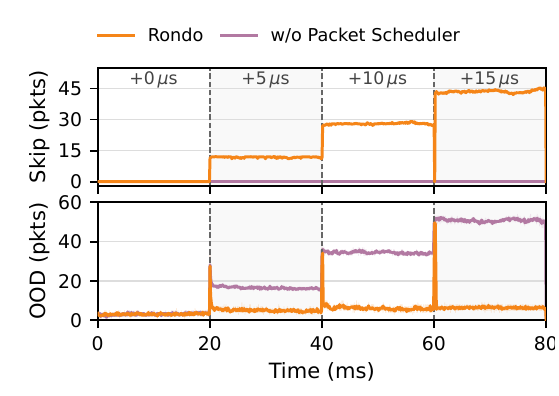}
        \caption{Packet scheduling reduces OOD.}
        \label{fig:eval_scheduler}
    \end{minipage}\hfill \begin{minipage}[t]{0.235\textwidth}
        \vspace{0pt}
        \centering
        \includegraphics[
            width=\linewidth
        ]{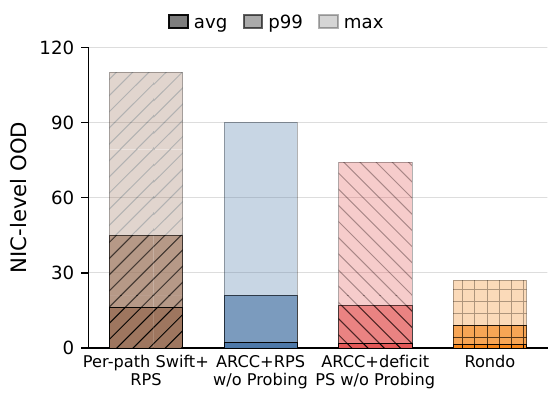}
        \caption{Ablation study of \sysname's components.}
        \label{fig:eval_ablation}
    \end{minipage}\hfill \begin{minipage}[t]{0.235\textwidth}
        \vspace{0pt}
        \centering
        \includegraphics[
            width=\linewidth
        ]{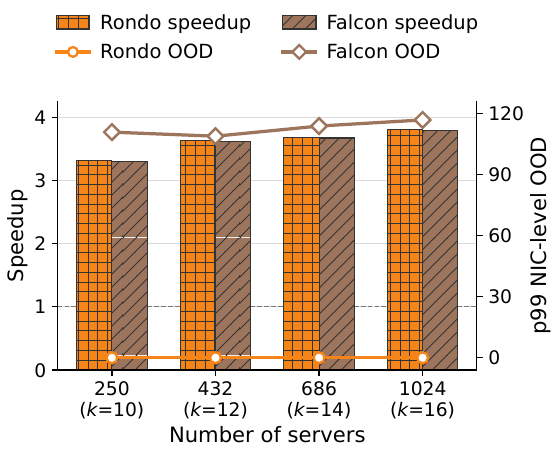}
        \caption{Scalability results.}
        \label{fig:eval_scalability}
    \end{minipage}\hfill \begin{minipage}[t]{0.235\textwidth}
        \vspace{0pt}
        \centering
        \includegraphics[
            width=\linewidth
        ]{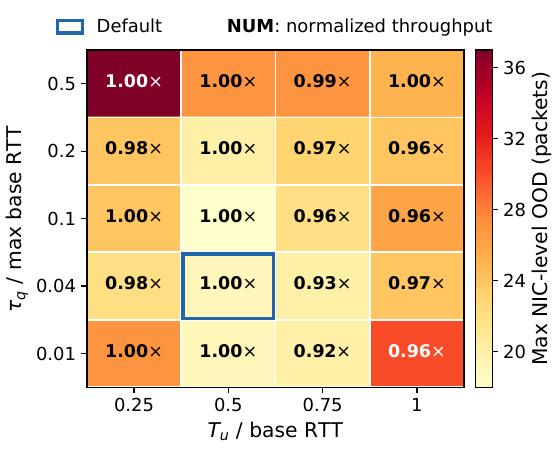}
        \caption{Parameter sweep.}
        \label{fig:eval_sensitivity}
    \end{minipage}
\end{figure*}

\para{Packet scheduler controls OOD under heterogeneity}
Figure~\ref{fig:eval_scheduler} illustrates how packet scheduling works.
We dynamically change the delay of one link while a flow is transmitting.
Without the scheduler, \sysname can maintain a stable queue,
but it cannot handle the persistent delay difference, causing OOD to increase.
The scheduler detects this difference after the delay changes and precisely adjusts the transmission order,
thereby maintaining a low OOD.

\begin{table*}[!t]
    \centering
    \small
    \begin{tabularx}{\textwidth}{
        @{}
        l
        >{\raggedright\arraybackslash}X
        >{\raggedright\arraybackslash}X
        >{\raggedright\arraybackslash}X
        >{\raggedright\arraybackslash}X
        @{}
    }
        \toprule
        \textbf{Transport}
        & \textbf{Congestion control}
        & \textbf{Path selection}
        & \textbf{Switch support}
        & \textbf{In-order delivery$^{1}$} \\
        \midrule

        RoCE AR~\cite{khashabHighspeedNetworkingGigaScale2026}
        & DCQCN or ZTR-RTT
        & Switch-driven AR
        & NVIDIA Spectrum
        & No \\

        SRD~\cite{shalevCloudOptimizedTransportProtocol2020}
        & Delay-based
        & Delay-based
        & Commodity Ethernet
        & No \\

        Falcon~\cite{singhviFalconReliableLow2025}
        & Per-path delay-based
        & Maximum-open-window
        & Commodity Ethernet
        & Optional$^{2}$ \\

        MP-RDMA~\cite{luMultiPathTransportRDMA2018}
        & Flow-level ECN-based
        & ACK-clocked, OOD-aware
        & Commodity Ethernet
        & No \\

        UET~\cite{ultraethernetconsortiumUltraEtherneSpecification2025}
        & Flow-level ECN + delay
        & ACK/ECN-based
        & Commodity Ethernet
        & Optional (single-path)$^{2}$ \\

        MRC~\cite{araujoResilientAISupercomputer}
        & Flow-level ECN + delay
        & ACK/ECN-based
        & Commodity Ethernet
        & No \\
        EQDS/NDP~\cite{olteanuEdgequeuedDatagramService2022,handleyRearchitectingDatacenterNetworks2017}
        & Receiver-driven
        & Random packet spray
        & Packet trimming
        & Yes$^{2}$ \\

        \midrule

        \textbf{\sysname}
        & Per-path ACK-rate-based CC with probing
        & Deficit-based
        & Commodity Ethernet
        & Yes$^{2}$ (controllable reorder-buffer demand) \\

        \bottomrule
        \multicolumn{5}{@{}l}{\footnotesize
        $^{1}$ In-order delivery refers to whether packets are delivered to the upper layer in order despite out-of-order arrival.} \\
        \multicolumn{5}{@{}l}{\footnotesize
        $^{2}$ UET provides in-order delivery only in single-path mode; the other transports use a reorder buffer.}
    \end{tabularx}
    \caption{Overview of representative datacenter packet spraying transports.}
    \label{tab:related-transport}
\end{table*}

\para{Ablation study}
Figure~\ref{fig:eval_ablation} shows the contribution of each \sysname component
under permutation traffic.
Replacing per-path Swift with ARCC reduces the average \oodname by 87.3\%
and the p99 \oodname by 53.3\%, because ARCC maintains stable queues
after convergence.
Deficit-based path selection further reduces OOD by smoothing packet assignments relative to random packet spray (RPS).
Capacity probing then significantly reduces the maximum and p99 OOD by avoiding queue overshoot during flow startup
and rapid rate increases.
The packet scheduler helps under persistent path heterogeneity, as shown in Figure~\ref{fig:eval_scheduler}.

\para{Scalability}
To evaluate the topology scalability of \sysname,
we measure its speedup relative to ECMP for all-to-all traffic on a $k$-ary fat-tree topology,
where $k$ ranges from 10 to 16.
Figure~\ref{fig:eval_scalability} shows that \sysname's speedup increases from 3.31$\times$ to 3.81$\times$,
while its p99 OOD remains zero packets.

\para{Parameter sensitivity}
Figure~\ref{fig:eval_sensitivity} shows that \sysname is robust to
$T_u$ and $\tau_q$, maintaining 0.92--1.00$\times$ normalized throughput
and a maximum OOD of 18--36 packets across all settings under permutation traffic.
A smaller $T_u$ enables faster reactions but
makes the ACK-rate measurements less stable.
A larger $\tau_q$ maintains a larger queue which
immediately utilizes newly available capacity
but generally increases OOD.

Additional evaluations are provided in Appendix~\ref{app:addeval}.

%% file: sections/conclusion.tex
\section{Related Work}

\para{Packet-level multipathing}
Datacenter transports~\cite{luMultiPathTransportRDMA2018,shalevCloudOptimizedTransportProtocol2020,ultraethernetconsortiumUltraEtherneSpecification2025,araujoResilientAISupercomputer,singhviFalconReliableLow2025,handleyRearchitectingDatacenterNetworks2017} and load-balancing schemes~\cite{dixitImpactPacketSpraying2013,bonatoREPSRecycledEntropy2026a,ghorbaniDRILLMicroLoad2017,qiSGLBScalableRobust2025,leSTrackReliableMultipath2024,liPSNPATHWhenMultipath2026,EnablingPacketSpraying,huFLBFinegrainedLoad2025} exploit path diversity through packet-level multipathing.
Table~\ref{tab:related-transport} summarizes representative transports.
These designs either do not support in-order delivery, fall back to single-path, or prioritize bandwidth utilization that may introduce significant packet reordering.
In contrast, \sysname fully utilizes all available paths while keeping OOD under control.

\para{Other methods to reduce reordering}
Conweave~\cite{songNetworkLoadBalancing2023} reorders packets at the ToR switch, but limited switch buffering constrains its scalability and path-switching granularity.
QDAPS~\cite{huangQDAPSQueueingDelay2018} and FLB~\cite{huFLBFinegrainedLoad2025} select paths based on queueing delays to mitigate reordering, but tend to direct subsequent packets onto a high-delay path to follow an earlier packet.
MP-RDMA~\cite{luMultiPathTransportRDMA2018} prunes paths with late ACKs to reduce reordering, leaving slower paths underutilized. DAPS~\cite{kuhnDAPSIntelligentDelayaware2014a} and STMS~\cite{shiSTMSImprovingMPTCP2018} are MPTCP~\cite{wischikDesignImplementationEvaluation2011} schedulers that reorder transmissions across subflows, but target low-rate Internet settings and rely on heuristic or reactive schemes that cannot precisely control OOD.
\sysname co-designs the multipath controller and packet scheduler for high-throughput datacenter transport.

\para{Congestion control}
ACC~\cite{zhangRevisitingCongestionControl2024} uses ACK rates to determine the pause duration after congestion, but relies on tentative rate increases to reclaim spare bandwidth, resulting in slow ramp-up and queue oscillations.
Aeolus~\cite{huAeolusBuildingBlock2020a} exploits spare bandwidth through speculative data transmission, but the  packets still interfere with normal data traffic and are more likely to be dropped.
\sysname instead uses ACK rates as capacity indicators and dedicated packets for probing to maintain stable queues and limit reordering.

A more comprehensive discussion of datacenter load balancing and congestion control is provided in Appendix~\ref{app:relatedwork}.

\section{Conclusion}
We present \sysname, a multipath transport that controls packet reordering while sustaining high throughput.
\sysname co-designs the multipath controller and packet scheduler: the controller limits transient queue differences across paths, while the scheduler compensates for persistent inter-path delay differences.
Together, these mechanisms show that high-throughput packet-level multipathing need not incur extensive reordering, providing a practical design point for transports operating with constrained NIC resources.

%% file: sections/appendix/appendix.tex
\input{sections/appendix/app_cx7ood.tex}
\input{sections/appendix/ooo_analysis.tex}
\input{sections/appendix/design_proof.tex}

\input{sections/appendix/app_implement.tex}
\input{sections/appendix/app_eval.tex}
\input{sections/appendix/app_relatedwork.tex}

%% file: sections/appendix/app_cx7ood.tex
\section{Resource limits in ConnectX-7}
\label{app:hw_cx7}

To further explore the resource limits in RNICs supporting DDP~\cite{OutoforderDataPlacement2026}, we measure two hosts equipped with ConnectX-7 NICs and connected through a Tofino2 switch over 40-Gbps links.
ConnectX-7 directly places out-of-order payloads into host memory and thus avoids an on-NIC payload buffer.
The switch routes selected RoCEv2 packets through artificially delayed paths to vary OOD.
We enable Selective Repeat and drop NAKs at the switch, allowing the receiver to accept out-of-order packets.\footnote{ConnectX-7 requires an NVIDIA Spectrum switch to enable adaptive routing, which supports out-of-order reception without NAKs.
Dropping NAKs at the switch approximates this behavior in our testbed.}

\begin{figure}
    \centering
    \includegraphics[width=0.8\linewidth]{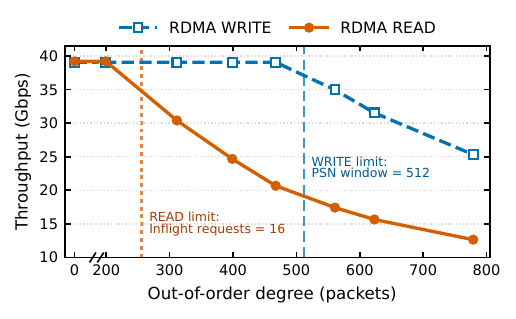}
    \caption{ConnectX-7 throughput degrades under excessive reordering.}
    \label{fig:cx7throughput}
\end{figure}

Figure~\ref{fig:cx7throughput} shows that RDMA READ throughput drops sharply when OOD exceeds approximately 200 packets; RDMA WRITE exhibits a similar cliff near 500 packets.
This occurs because ConnectX-7 limits each queue pair to 16 outstanding READ operations and a 512-packet outstanding PSN range for WRITE~\cite{nvidia_cx6dx_fw_release_notes,nvidia_kernel_upstream_release_notes,liScaleAcrossExplorerExploring2026}.
Once either limit is reached, the NIC stalls further transmission.
It exhibits the metadata resource constraint for an RNIC.

Moreover, the tolerance to reordering also depends on the network BDP.
The outstanding PSN window for RDMA WRITE must accommodate both normal inflight packets and reordered packets.
Therefore, in high-speed datacenter networks with large BDPs, a substantial portion of this window is consumed by normal inflight, leaving less room for packet reordering.
Our testbed uses a smaller BDP (40 Gbps $\times$ 10 $\mu$s), allowing the outstanding window to be primarily used for absorbing reordering.
However, network with a 200-Gbps line rate and a 40-$\mu$s RTT has a BDP of approximately 244 packets, consuming nearly half of the 512-packet PSN window before accounting for reordering.

%% file: sections/appendix/ooo_analysis.tex
\section{Analysis of OOD and RBO}
\label{app:general-ood-analysis}

We derive the OOD, the maximum OOD estimate, and the
corresponding RBO used in \S\ref{sec:understanding}.
Packets are transmitted in increasing PSN order.
All sending rates below are measured in
packets per unit time.

\para{OOD derivation}
Consider packet $i$, transmitted at time $t_i$ on path $p_i$, that is
the next expected packet immediately before its arrival.
A later packet $j$ overtakes it if
\begin{align}
    t_j > t_i
    \quad\text{and}\quad
    t_j + D_{p_j}(t_j) < t_i + D_{p_i}(t_i)
    \label{eq:overtake-condition-delay}
\end{align}
Let $j$ be the highest-PSN packet that overtakes packet $i$.
Immediately before packet $i$ arrives, OOD equals $j-i$.
Since PSNs increase in transmission order, this gap counts all packets
transmitted during $(t_i,t_j]$, including those that have not yet arrived.
Using the aggregate sending rate $R(u)$ gives
\begin{align}
    OOD = j-i \approx \int_{t_i}^{t_j} R(u)\,du
    \label{eq:general-ood-integral}
\end{align}
This yields Equation~\eqref{eq:psn_gap} in the main text.

\para{Maximum OOD estimation}
Assume path delays remain approximately constant over the overtaking
window, and denote the maximum inter-path delay difference by
\begin{align}
    \Delta D(t)
    = \max_{p\in\mathcal{P}} D_p(t)
    - \min_{p\in\mathcal{P}} D_p(t)
    \label{eq:appendix-delay-spread}
\end{align}
To obtain the maximum overtaking window, consider packet $i$ transmitted
on a maximum-delay path.
The overtaking condition then requires $t_j-t_i<\Delta D(t)$.
Let $g$ be the maximum gap between consecutive transmissions on a
minimum-delay path over this interval.
When $g<\Delta D(t)$, a packet on that path is transmitted within $g$
before $t_i+\Delta D(t)$ and arrives before packet $i$.
Thus, the maximum overtaking window satisfies
\begin{align}
    \Delta D(t)-g
    &\le \max_{j\text{ overtakes }i}(t_j-t_i)
    \nonumber\\
    &< \Delta D(t)
    \label{eq:inter_path_delay_diff}
\end{align}
In high-throughput transport, $g$ is negligible relative to
$\Delta D(t)$, so the maximum overtaking window is
approximately $\Delta D(t)$.
Letting $\bar R$ denote the average aggregate sending rate over the
window, Equation~\eqref{eq:general-ood-integral} gives
\begin{align}
    OOD_{\max}
    &\approx \bar R\Delta D(t)
    \nonumber\\
    &= \bar R
    \left(
        \max_{p\in\mathcal{P}} D_p(t)
        - \min_{p\in\mathcal{P}} D_p(t)
    \right)
    \label{eq:appendix-ood-estimation}
\end{align}
This is the estimate in Equation~\eqref{eq:ood_estimation}.

\para{RBO derivation}
Consider packet $i$ that is the next expected packet at the receiver
immediately before its arrival. At that instant, RBO counts the
later packets that have already arrived and are buffered waiting for
packet $i$.
Consider steady-state paths with constant forward delays $D_p$ and
sending rates $r_p$.
A later packet $j$, transmitted at time $t_j>t_i$ on path $p_j$,
overtakes packet $i$ if
\begin{align}
    t_j+D_{p_j} < t_i+D_{p_i}
    \quad\Longleftrightarrow\quad
    t_j-t_i < D_{p_i}-D_{p_j}
    \label{eq:steady-overtake-condition}
\end{align}
The overtaking window on path $p$ therefore has length
$(D_{p_i}-D_p)^+$, where $(x)^+=\max(x,0)$.
The RBO immediately before packet $i$ arrives is the number of packets
that overtake it. Approximating this count by integrating the sending
rate over each path's window gives
\begin{align}
    RBO
    &=
    \left|\left\{
        j>i : t_j+D_{p_j}<t_i+D_{p_i}
    \right\}\right|
    \nonumber\\
    &\approx
    \sum_{\substack{p\in\mathcal{P}\\p\neq p_i}}
    \int_{t_i}^{t_i+(D_{p_i}-D_p)^+} r_p\,du
    \nonumber\\
    &=
    \sum_{\substack{p\in\mathcal{P}\\p\neq p_i}}
    r_p(D_{p_i}-D_p)^+
    \label{eq:steady-rbo}
\end{align}
These packets form a subset of the PSN span counted by OOD, so
$RBO\le OOD$.
It also motivates our packet scheduler design.

%% file: sections/appendix/design_proof.tex
\section{Proofs of the Propositions}
\label{app:theory_analysis}

\subsection{Model and Assumptions}
\label{app:arcc_model}

Consider a bottleneck of capacity $C$.
We assume each competing flow has one logical path mapped
to the physical path passing this bottleneck.
In practice, if a flow has multiple logical paths
mapped to it, this proof treats them as single flows
because ARCC controls per-path windows.
We assume an ideal fluid model with a single bottleneck.
We write $W_f$, $q_f$, $Q_f$, $\hat r_f$, and
$T_f$ for the corresponding per-path variables of flow $f$
defined in \S\ref{sec:arcc}.
Since ARCC updates windows at fixed intervals, we use index
k to represent the state after the k-th update, e.g., $W_f[k]$.
Flows that share the same bottleneck experience the same queueing delay $d$.
Let $I_f$ denote \texttt{pktInfl}, and use $a$ for the
additive increase step size as $a=\frac{q_0}{4}$.

\subsection{Proof of Fairness}
\label{app:fairness_proof}

\begin{proof}[Proof of Proposition~\ref{prop:arcc_fairness}]
Both probing and fast-correction are temporary states
for ARCC when the set of flows remains unchanged,
and its operation uses the EWMA window update described in \S\ref{sec:arcc}:
\begin{align}
    W_f[k+1]=(1-\gamma)W_f[k] + \gamma (T_f\hat r_f[k]+q_f[k+1])
    \label{eq:window_update_algorithm}
\end{align}
During these normal control epochs, the inflight matches the window fast, and we have
\begin{align}
    W_f[k]=(T_f+d[k])\hat r_f[k]
    \label{eq:window_rate_realationship}
\end{align}
We assume every $H$ consecutive updates contain at least one
additive-increase update, and flows that share the same bottleneck observe
the same queueing delay.

\para{Margin converges}
Define the normalized disagreement as
\begin{align}
    \Phi[k]
    =1-\frac{\min_f q_f[k]}{\max_f q_f[k]}
    \label{eq:normalized_margin_disagreement}
\end{align}

All contenders use
the same multiplicative-decrease factor when $d[k]>\tau_q$:
\begin{equation}
    q_f[k+1]
    =\frac{\tau_q}{d[k]}q_f[k]
\end{equation}
Such an update leaves $\Phi$ unchanged.

For an additive-increase update
$q_f[k+1] = \min\{q_0, q+a\}$,
for $0<v\le u\le q_0$, we have
\begin{equation}
    1-\frac{\min\{q_0,v+a\}}{\min\{q_0,u+a\}}
    \le
    \left(1-\frac{a}{q_0}\right)
    \left(1-\frac{v}{u}\right)
\end{equation}
Since $a=q_0/4$, every additive-increase update contracts
$\Phi$ by at least a factor of $3/4$.
Therefore,
\begin{equation}
 \Phi[k]
    \le
    \left(\frac34\right)^{\lfloor k/H\rfloor}\Phi[0]
\label{eq:margin_geometric_convergence}
\end{equation}
It means that the margins converge to fairness, and the
margin disagreement decreases geometrically.

\para{Window and rate dynamics}
Under the relation in Equation~\eqref{eq:window_rate_realationship},
the window update gives
\begin{align}
        &(T_f+d[k+1])\hat r_f[k+1] \\
        &\quad=
        \bigl(T_f+(1-\gamma)d[k]\bigr)\hat r_f[k]
        +\gamma q_f[k+1]
    \label{eq:joint_window_rate_update}
\end{align}

Since
$\sum_f\hat r_f[k+1]=\sum_f\hat r_f[k]=C$,
summing Equation~\eqref{eq:joint_window_rate_update} yields
\begin{equation}
    d[k+1]-(1-\gamma)d[k]
    =
    \gamma
    \frac{
        \displaystyle\sum_f
        \frac{q_f[k+1]}{T_f+d[k+1]}
    }{
        \displaystyle\sum_f
        \frac{\hat r_f[k]}{T_f+d[k+1]}
    }
    \label{eq:joint_delay_increment}
\end{equation}

Define the weighted averages
\begin{align}
    \bar r[k]
    &=
    \frac{
        \displaystyle\sum_f
        \frac{\hat r_f[k]}{T_f+d[k+1]}
    }{
        \displaystyle\sum_f
        \frac{1}{T_f+d[k+1]}
    },
    \bar q[k+1]
    &=
    \frac{
        \displaystyle\sum_f
        \frac{q_f[k+1]}{T_f+d[k+1]}
    }{
        \displaystyle\sum_f
        \frac{1}{T_f+d[k+1]}
    }
\end{align}
Equation~\eqref{eq:joint_delay_increment} then becomes
\begin{align}
    d[k+1]-(1-\gamma)d[k]
    =
    \gamma\frac{\bar q[k+1]}{\bar r[k]}
\end{align}

Substituting this identity into
Equation~\eqref{eq:joint_window_rate_update} gives
\begin{align}
        \hat r_f[k+1]=\hat r_f[k] +
        \frac{d[k+1]-(1-\gamma)d[k]}
             {T_f+d[k+1]} (\frac{q_f[k+1]\bar r[k]}{\bar q[k+1]}-\hat r_f[k])
    \label{eq:rate_mixing_update}
\end{align}

Define the coefficient
\begin{equation}
        \eta_f[k]
        =
        \frac{d[k+1]-(1-\gamma)d[k]}
             {T_f+d[k+1]}
\end{equation}
Equation~\eqref{eq:rate_mixing_update} can then be
written as
\begin{align}
        \hat r_f[k+1]
        =
        (1-\eta_f[k])\hat r_f[k]
        +
        \eta_f[k]
        \frac{q_f[k+1]}{\bar q[k+1]}\bar r[k]
    \label{eq:rate_mixing_update_final}
\end{align}
This update is a convex combination
of the current estimated rate and the target
$\frac{q_f[k+1]}{\bar q[k+1]} \bar r[k]$.
When the margins $q_f$ are equal, every flow moves toward $\bar r[k]$.

\para{Convergence rate}
The mixing coefficients satisfy
\begin{equation}
    0<
    \eta
    \triangleq
    \frac{\gamma q_-}{C(\max_g T_g+d_+)}
    \le
    \eta_f[k]
    <1
    \label{eq:rate_mixing_lower_bound}
\end{equation}
where $q_->0$ and $d_+<\infty$ denote a lower bound on $q_f[k]$
and an upper bound on $d[k]$, respectively.
The former follows from the bounded interval between additive
increases, and the second one from the network queue bound.

If all margins have converged to the same value,
Equation~\eqref{eq:rate_mixing_update_final} moves every
$\hat r_f[k]$ toward the same weighted mean $\bar r[k]$.
This update contracts the rate range by a factor of at most
$1-\eta$.

For unequal margins, the deviation from this common target
satisfies
\begin{equation}
    \left|
        \frac{q_f[k+1]}{\bar q[k+1]}-1
    \right|
    \le
    \frac{q_0}{q_-}\Phi[k+1]
\end{equation}
Consequently,
\begin{equation}
    \begin{aligned}
        &\max_f\hat r_f[k+1]-\min_f\hat r_f[k+1] \\
        &\quad\le
        (1-\eta)
        \left(
            \max_f\hat r_f[k]-\min_f\hat r_f[k]
        \right) +
        \frac{2Cq_0}{q_-}\Phi[k+1]
    \end{aligned}
    \label{eq:joint_fairness_contraction}
\end{equation}

Combining Equations~\eqref{eq:margin_geometric_convergence}
and~\eqref{eq:joint_fairness_contraction} yields
\begin{equation}
    \max_f\hat r_f[k]-\min_f\hat r_f[k]
    =
    O(\rho^k)
\end{equation}
where $\rho$ satisfies
\begin{equation}
    \rho=\max\left\{
        1-\eta,
        \left(\frac34\right)^{1/H}
    \right\}
    \label{eq:fairness_convergence_factor}
\end{equation}
Because $\sum_f\hat r_f[k]=C$, every flow's measured ACK rate
therefore converges geometrically to $C/N$,
i.e., every flow's throughput converges geometrically to $C/N$.
\end{proof}

\subsection{Proof of the Out-of-Order Degree Bound}
\label{app:ood_proof}

\begin{proof}[Proof of Proposition~\ref{prop:ood_bound}]
We consider a bottleneck with at most $N_l$ controlled flows
and an isolated batch of at most $N_S$ starting flows.
Let $Q[k]=Cd[k]$ denote its physical queue occupancy.
Define $T_{\min}=\min_f T_f$ and $T_{\max}=\max_f T_f$.
We assume $\gamma \ge 1/2$.

\para{Queue bound with controlled flows}
We first consider a control phase with a fixed set
of at most $N_l$ flows and use $Q[k]$ to represent the
bottleneck queue after the k-th update.
Flows can leave but do not join in this phase.
We assume the inflight matches the window fast enough.
Combining equation~\eqref{eq:window_rate_realationship} and
the window update algorithms, we get
\begin{align}
    Q[k] &= \sum_f (W_f[k] -  T_f\hat r_f[k]) \\
    Q[k+1] &= (1-\gamma)Q[k]
    +\gamma \sum_f q_f[k+1] + \sum_f T_f(\hat r_f[k] -\hat r_f[k+1])
\end{align}

When the bottleneck queue is empty, the queue bound holds trivially.
We therefore consider epochs with a nonempty queue, during which the
work-conserving bottleneck is fully utilized and
$\sum_f \hat r_f[k]=C$.
Define $U[k]$ as follows and combine it with equation~\eqref{eq:window_rate_realationship}:
\begin{equation}
    U[k]
    \triangleq
    \sum_f\frac{W_f[k]}{T_f}-C
    =
    \frac{Q[k]}{C}
    \sum_f\frac{\hat r_f[k]}{T_f}
    \label{eq:weighted_excess}
\end{equation}
We have
\begin{equation}
    U[k]T_{\min} \le Q[k]
    \le U[k]T_{\max}
    \label{eq:weighted_excess_queue}
\end{equation}
and
\begin{equation}
    U[k+1]
    =
    (1-\gamma)U[k]
    +
    \gamma\sum_f\frac{q_f[k+1]}{T_f}
    \label{eq:weighted_excess_update}
\end{equation}

Each update increases $q_f$ by at most $a$, so
\begin{equation}
    \begin{aligned}
        &\sum_f\frac{q_f[k+1]}{T_f}-U[k+1] \le
        (1-\gamma)
        \left(
            \sum_f\frac{q_f[k]}{T_f}-U[k]
            +\frac{N_l a}{T_{\min}}
        \right)
    \end{aligned}
    \label{eq:weighted_lag_update}
\end{equation}
The margin is initialized as $q_f=a=q_0/4$, and the first update uses the direct window correction described in \S\ref{sec:arcc}. After the first update
\begin{align}
    \sum_f\frac{q_f[1]}{T_f}-U[1] &= 0
    \le
    \frac{1-\gamma}{\gamma}
    \frac{N_l a}{T_{\min}}
    \label{eq:weighted_lag_initial} \\
U[1]&\le \frac{C\tau_q+N_l a}{T_{\min}}
\end{align}
By mathematical induction,
\begin{equation}
    \sum_f\frac{q_f[k]}{T_f}-U[k]
    \le
    \frac{1-\gamma}{\gamma}
    \frac{N_l a}{T_{\min}}
    \label{eq:weighted_lag_bound}
\end{equation}

When $U[k]\le C\tau_q/T_{\min}$, Equations
\eqref{eq:weighted_excess_update} and
\eqref{eq:weighted_lag_bound} give
\begin{equation}
    \begin{aligned}
        U[k+1]
        &\le
        U[k]+\gamma
        \left(
            \frac{1-\gamma}{\gamma}
            \frac{N_l a}{T_{\min}}
            +\frac{N_l a}{T_{\min}}
        \right) \\
        &\le
        \frac{C\tau_q+N_l a}{T_{\min}}
    \end{aligned}
    \label{eq:weighted_ai_bound}
\end{equation}

When $U[k]>C\tau_q/T_{\min}$,
Equation~\eqref{eq:weighted_excess_queue} implies
$Q[k]>C\tau_q$, so all flows perform MD.
Moreover,
\begin{equation}
    \begin{aligned}
        \sum_f\frac{q_f[k+1]}{T_f}
        &=
        \frac{C\tau_q}{Q[k]}
        \sum_f\frac{q_f[k]}{T_f} \\
        &\le
        \frac{C\tau_q}{T_{\min}U[k]}
        \left(
            U[k]
            +\frac{1-\gamma}{\gamma}
             \frac{N_l a}{T_{\min}}
        \right)
    \end{aligned}
\end{equation}
Substituting into
Equation~\eqref{eq:weighted_excess_update} yields
\begin{equation}
    U[k+1]
    \le
    (1-\gamma)U[k]
    +\frac{\gamma C\tau_q}{T_{\min}}
    +\frac{(1-\gamma)C\tau_q N_l a}
           {T_{\min}^2 U[k]}
    \label{eq:weighted_md_bound}
\end{equation}
The right-hand side is convex in $U[k]$.
For $\gamma\ge1/2$, its values on the interval
$
    \left[
        \frac{C\tau_q}{T_{\min}},
        \frac{C\tau_q+N_l a}{T_{\min}}
    \right]
$
are at most $(C\tau_q+N_l a)/T_{\min}$.
Above this interval, it is no larger than $U[k]$.
Combining the two cases therefore gives
\begin{equation}
    U[k]
    \le
        \frac{C\tau_q+N_l a}{T_{\min}}
    \label{eq:weighted_excess_bound}
\end{equation}

Then
\begin{equation}
    Q[k]
    \le
    \frac{T_{\max}}{T_{\min}}
    \left(C\tau_q+N_l a\right)
    \label{eq:normal_queue_bound}
\end{equation}

\para{Queue bound when flows join}
Adding at most $N_S$ flows' initial windows increases $U$ by at
most $N_SW_0/T_{\min}$.
Thus, immediately after their admission,
\begin{equation}
    U
    \le
    \frac{C\tau_q+N_l a+N_SW_0}{T_{\min}}
    \label{eq:startup_weighted_excess}
\end{equation}

Applying Equation~\eqref{eq:weighted_excess_bound} to the
post-startup stages gives
\begin{equation}
    \begin{aligned}
        U[k]
        &\le
        \max\left\{
            \frac{C\tau_q+N_l a+N_SW_0}{T_{\min}},
            \frac{C\tau_q+(N_l+N_S)a}{T_{\min}}
        \right\} \\
        &=
        \frac{
            C\tau_q+N_l a+N_SW_0
        }{T_{\min}}
    \end{aligned}
\end{equation}
Since $W_0\ge a=q_0/4$. Therefore, throughout the modeled startup episode,
\begin{equation}
    Q[k]
    \le
    \frac{T_{\max}}{T_{\min}}
    \left(
        C\tau_q+N_l a+N_SW_0
    \right)
    \label{eq:combined_queue_bound}
\end{equation}

\para{OOD bound}
We finally translate the queue bound into a bound on the OOD caused by
inter-path queueing-delay differences. We assume that each path
traverses at most one congested bottleneck of capacity $C$.
For any considered flow, its aggregate sending rate is at most
$R_{\max}$. Therefore,
the upper bound on the \oodname is
\begin{equation}
    OOD
    \le
    \frac{R_{\max}}{C}
        \frac{T_{\max}}{T_{\min}}
        \left(
            C\tau_q
            +\frac{N_l q_0}{4}
            +N_S W_0
        \right).
    \label{eq:controller_ood_bound}
\end{equation}
This is a conservative worst-case bound.
Equality is achieved only when one path of a flow reaches the maximum queue bound
while another path has zero queue. As the number of competing flows increases,
this worst-case condition becomes less likely to occur.
Meanwhile, with more competing flows, the likelihood that an individual
flow transmits at a high rate also decreases.
\end{proof}

\input{sections/appendix/app_prosition3proof}

%% file: sections/appendix/app_prosition3proof.tex
\subsection{Proof of the Packet Scheduling Correctness}
\begin{proof}[Proof of Proposition~\ref{prop:delay_compensation}]
Consider all transmission opportunities of a flow before transmission
starts. Since each path has a fixed sending rate and forward delay, the
transmission time, path, and arrival time of every opportunity are
known. Consider an ideal scheduler that sorts these opportunities by
their arrival times and assigns consecutive PSNs in this order. This
defines a permutation of the PSNs under which packets arrive in order.

We show that \sysname reproduces this permutation online. Consider the
\(k\)-th transmission opportunity, on path \(p\) at time \(t\). Its
global rank in the ideal scheduler is one plus the number of all
opportunities that arrive before it.

By the \(k\)-th transmission, the first \(k-1\) opportunities have
already consumed their assigned PSNs. Among them, only those that arrive
before the current opportunity have smaller PSNs in the ideal
assignment. Therefore, the rank of the current opportunity among the
remaining unsent PSNs is one plus the number of future
opportunities that arrive before it.

A future opportunity on path \(p'\), transmitted at \(t'>t\), arrives
before the current one if
\begin{equation}
    t' + D_{p'} < t + D_p
\end{equation}
or equivalently
\begin{equation}
    t'-t < D_p-D_{p'}
\end{equation}

Under the fluid model, path \(p'\) contributes
\(r_{p'}(D_p-D_{p'})^+\) such opportunities. Thus, after rounding the
fluid-model result to an integer packet count, the rank of the current
opportunity among the remaining opportunities is
\begin{equation}
    1+
    \left\lfloor
    \sum_{\substack{p'\in\mathcal P\\p'\neq p}}
    r_{p'}(D_p-D_{p'})^+
    \right\rfloor
\end{equation}

This is exactly the \((skip_p+1)\)-th smallest unsent PSN selected by
\sysname. Therefore, whenever all previous transmission opportunities
match the ideal permutation, the current one also receives the same PSN.
Starting from the first transmission opportunity, induction shows that
\sysname reproduces the entire ideal permutation online. Hence, packets
arrive in increasing PSN order and
\begin{equation}
    OOD = 0
\end{equation}
\end{proof}

%% file: sections/appendix/app_implement.tex
\section{Implementation Details and Optimizations}
\label{app:imple_detail}
\para{Implementation details}
In our DPDK prototype, the basic multipath transport supports per-packet path selection,
per-packet ACKs, bitmap-based packet arrival tracking
and selective repeat.
TSC-based timers trigger ARCC updates at fixed intervals, which update
the per-path windows and the flow-level pacing rate.
For each transmission, deficit-based path selection selects the
logical path and then packet scheduling decides
the packet to transmit.
The transport changes the UDP source port field to
select the logical path for packets,
and transmits packets through \texttt{rte\_eth\_tx\_burst()}.
To avoid reverse-path congestion from distorting ACK-rate measurement,  ACK packets use the highest priority.

\para{Overhead optimizations}
We estimate arithmetic overhead based
on~\cite{OverviewSifive7seriesMicroarchitecture},
using a cost model that assigns 1 cycle to an integer addition or subtraction, 3 cycles to
an integer multiplication, and 16 cycles to an integer division.
We apply two optimizations to reduce this overhead.
First, we separate per-packet operations from periodic control operations.
ARCC window updates and per-path skip calculations are performed
periodically off the critical path.
Second, divisions by $T$ and $T_u$ are replaced with multiplications by
precomputed reciprocals, and the normalization division in deficit-based
path selection is replaced with a shared reciprocal table.
After these optimizations, for $P$ paths, path selection requires $6P$ arithmetic cycles, and selecting the target packet using a
precomputed skip requires 1 cycle.
Every $T_u$, ARCC requires approximately $(35\text{--}45)P$ cycles, while
precomputing the per-path skips requires at most $\frac{5P(P-1)}{2}$ cycles.

%% file: sections/appendix/app_eval.tex
\section{Additional Evaluation}
\label{app:addeval}
\begin{figure*}[t]
    \centering
    \includegraphics[width=\linewidth]{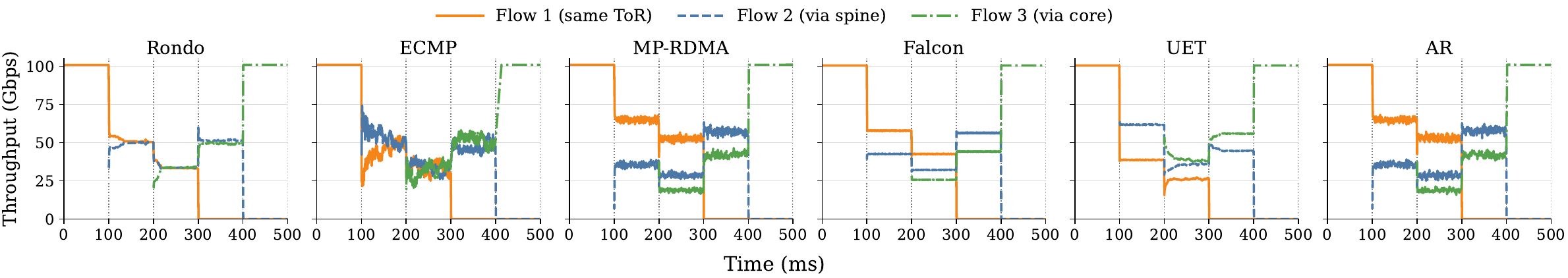}
    \caption{Throughput curves when three flows with different base RTTs step in and out.}
    \label{fig:fairness_3flows}
\end{figure*}

\begin{figure}[t]
    \centering
    \begin{minipage}[t]{0.53\linewidth}
        \centering
        \includegraphics[width=\linewidth]{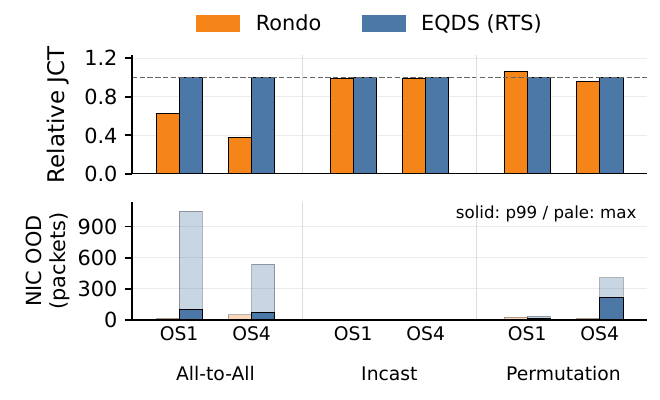}
        \caption{
Comparison between \sysname and EQDS under typical traffic.
        }
        \label{fig:eval_eqds}
    \end{minipage}
    \hfill
    \begin{minipage}[t]{0.45\linewidth}
    \centering
    \includegraphics[width=\linewidth]{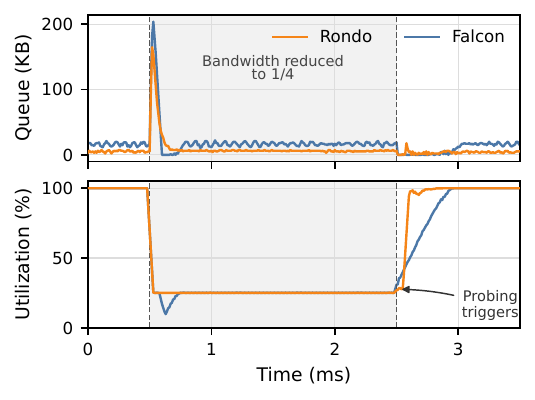}
    \caption{\sysname converges quickly when bandwidth changes.}
    \label{fig:eval_bwchanges}
    \end{minipage}
\end{figure}

\para{\sysname provides fairness for flows with different RTTs}
FRCC~\cite{agarwalFRCCProvablyFair2026} shows that fairness is difficult to achieve among flows with different base RTTs.
In DCNs, base RTT differences between flows are generally limited,
as they primarily stem from differences in path hop count.
Figure~\ref{fig:fairness_3flows} shows that \sysname keeps fairness among three flows sending to the same receiver from different locations.
One sender is in the same rack as the receiver, another is in the same pod, and the third is in a different pod.
This result is consistent with Proposition~\ref{prop:arcc_fairness}.

\para{Comparison to EQDS} EQDS~\cite{olteanuEdgequeuedDatagramService2022} is a receiver-driven transport,
where the sender transmits data packets only after receiving credits.
We compare \sysname with EQDS in Figure~\ref{fig:eval_eqds},
with EQDS implemented in RTS mode without packet trimming support.
EQDS controls OOD well under incast, but is less effective at handling
congestion in the fabric.
Therefore, its performance degrades under 4:1 oversubscription and all-to-all traffic.
In comparison, \sysname does not face these limitations.

\para{\sysname reacts quickly to bandwidth changes}
Figure~\ref{fig:eval_bwchanges} shows how \sysname reacts when the link bandwidth
drops to one quarter and later recovers.
There are 4 long flows in this experiment.
When the bandwidth decreases, transient queue buildup is unavoidable,
but ARCC quickly estimates the new capacity and drains the queue
without underutilizing the link.
When the bandwidth recovers, capacity probing enables rapid ramp-up
without overshooting.

%% file: sections/appendix/app_relatedwork.tex
\section{Discussions on Related Works}
\label{app:relatedwork}

\para{Congestion Control}
Sender-driven CCs~\cite{zhuCongestionControlLargeScale2015,liHPCCHighPrecision2019,addanki2022powertcp,kumarSwiftDelaySimple2020,arslanBoltSubRTTCongestion2023,zhangRevisitingCongestionControl2024,mittalTIMELYRTTbasedCongestion2015a,zhangoscar26,brakmo1994tcp,cardwell2016bbr} mainly focus on legacy single-path transport,
while \sysname is designed for multipath transport.
Receiver-driven congestion control schemes~\cite{montazeriHomaReceiverDrivenLowLatency2018a,olteanuEdgequeuedDatagramService2022,handleyRearchitectingDatacenterNetworks2017,huAeolusBuildingBlock2020a} mitigate performance degradation caused by incast but can
not directly capture congestion within the network fabric.
\sysname's ACK-based method
provides similar fine-grained control while capturing congestion in
both the fabric and the last hop.

ACC~\cite{zhangRevisitingCongestionControl2024} observes the ACK rate and uses it to determine the stall time after congestion, achieving fast convergence in congestion.
But it still relies on TCD~\cite{zhangRevisitingCongestionDetection2023} to infer congestion, tentatively increases its rate to utilize spare bandwidth, which results in slow rate increasing and queue oscillations.
\sysname uses ACK rate throughout its control loop and maintains stable queues.

OSCAR~\cite{zhangoscar26} achieves $\mathcal{O}(1)$ convergence for both rate increasing and reduction. However, it needs a sufficient queue   for measuring delay gradient, which introduces
a large inter-path queue difference in multipathing. This method also cannot work when flows start up.

Aeolus~\cite{huAeolusBuildingBlock2020a} allows senders to transmit
some low-priority data without credit during the first RTT for probing.
However, these packets use the same queue as normal packets, causing
a delay difference for existing flows.
It can obtain receiver credits, but cannot capture congestion information within the fabric before being dropped.
Moreover, these packets are more likely to be dropped, increasing the disorder.
The combination of probe packets and ACK-rate-based CC in \sysname addresses both problems.

\para{Datacenter load balancing}
Datacenter load-balancing schemes operate at different granularities.
Flow-level load-balancing methods~\cite{hoppsAnalysisEqualCostMultiPath2000,zhouWCMPWeightedCost2014,al-faresHederaDynamicFlow} cannot fully utilize the network with a small number of active flows.
Flowlet-level schemes~\cite{vaniniLetItFlow2017,qureshiPLBCongestionSignals2022,kabbaniFlowBenderFlowlevelAdaptive2014,alizadehCONGADistributedCongestionaware2014a} require sufficiently large inter-packet gaps to switch paths,
which are scarce in the bursty traffic of ML workloads.
Packet-level load balancing~\cite{dixitImpactPacketSpraying2013,bonatoREPSRecycledEntropy2026a,ghorbaniDRILLMicroLoad2017,qiSGLBScalableRobust2025,leSTrackReliableMultipath2024,liPSNPATHWhenMultipath2026,EnablingPacketSpraying,huFLBFinegrainedLoad2025} can better exploit path diversity,
but also introduces packet reordering.
Packet-level load balancing schemes can be further classified based on their congestion awareness.
Congestion-oblivious methods~\cite{dixitImpactPacketSpraying2013,liPSNPATHWhenMultipath2026,EnablingPacketSpraying} do not adapt traffic
allocation to path conditions and lead to load imbalance~\cite{qiSGLBScalableRobust2025}.
Locally congestion-aware methods~\cite{ghorbaniDRILLMicroLoad2017} make forwarding decisions using switch-local signals, enabling fast reactions
but with limited visibility of end-to-end path conditions.
End-to-end-congestion-aware methods~\cite{qiSGLBScalableRobust2025,leSTrackReliableMultipath2024,bonatoREPSRecycledEntropy2026a,singhviFalconReliableLow2025} make path-selection decisions at end hosts based on end-to-end congestion signals,
but still introduce substantial packet reordering.
\sysname performs packet-level multipathing based on end-to-end congestion information while keeping the resulting
reordering controllable.

MP-RDMA~\cite{luMultiPathTransportRDMA2018} introduces OOD-aware path selection, which
prunes paths with late ACKs to avoid excessive disorder.
But under a high inter-path queue difference, this method frequently prunes paths,
resulting in severe oscillations.
Even under a persistent delay difference,  it hurts utilization.

\para{Schedulers}
DAPS~\cite{kuhnDAPSIntelligentDelayaware2014a} constructs packet/path schedules heuristically from RTT
ratios and congestion windows.
However, its coarse schedule cannot precisely mitigate reordering.
It is limited to low-rate scenarios with stable congestion windows,
and its behavior perturbs the network state established by congestion control.
STMS~\cite{shiSTMSImprovingMPTCP2018} dynamically adapts sending packets allocated to different paths. It focuses on two-path scenarios, tentatively and reactively adjusts a transmission gap based on observed OOD.
It faces a trade-off between slow convergence and gap overshoot.